\documentclass[12pt,a4paper]{article}
\usepackage{amsmath,amssymb,graphicx,hyperref,slashed}
\usepackage{cite}
\numberwithin{equation}{section}
\allowdisplaybreaks[3]

\begin{document}
\begin{titlepage}
		
\renewcommand{\thefootnote}{\fnsymbol{footnote}}
\begin{flushright}
\begin{tabular}{l}
YITP-17-82\\
\end{tabular}
\end{flushright}
		
\vfill
\begin{center}
			
			
\noindent{\large \textbf{Three point functions in higher spin AdS$_3$ holography}}
			
\medskip
			
\noindent{\large \textbf{with $1/N$ corrections}}  
			
\vspace{1.5cm}

\noindent{Yasuaki Hikida$^{a}$\footnote{E-mail: yhikida@yukawa.kyoto-u.ac.jp} and Takahiro Uetoko$^b$\footnote{E-mail: rp0019fr@ed.ritsumei.ac.jp}}

\bigskip
			
\vskip .6 truecm
			
\centerline{\it $^a$Center for Gravitational Physics, Yukawa Institute for Theoretical Physics,}
\centerline{\it  Kyoto University, Kyoto 606-8502, Japan}
\medskip 
\centerline{\it $^b$Department of Physical Sciences, College of Science and Engineering,} 
\centerline{\it Ritsumeikan University, Shiga 525-8577, Japan}

\end{center}
		
\vfill
\vskip 0.5 truecm

\begin{abstract}

We examine three point functions with two scalar operators and a higher spin current in 2d W$_N$ minimal model to the next non-trivial order in $1/N$ expansion. The minimal model was proposed to be dual to a 3d higher spin gauge theory, and $1/N$ corrections should be interpreted as quantum effects in the dual gravity theory. We develop a simple and systematic method to obtain three point functions by decomposing four point functions of scalar operators with Virasoro conformal blocks. Applying the method, we reproduce known results at the leading order in $1/N$ and obtain new ones at the next leading order. As confirmation, we check that our results satisfy relations among three point functions conjectured before.
\end{abstract}
\vfill
\vskip 0.5 truecm
		
\setcounter{footnote}{0}
\renewcommand{\thefootnote}{\arabic{footnote}}
\end{titlepage}
	
\newpage
	
\tableofcontents

\section{Introduction}

Holography is expected to offer a way to learn quantum corrections of gravity theory from $1/N$ corrections in dual conformal field theory.
In this paper, we address this issue by utilizing one of the simplest holographies proposed in 
\cite{Gaberdiel:2010pz},%
\footnote{Recently, a different method to the issue has been adopted in \cite{Aharony:2016dwx,Alday:2017xua,Aprile:2017bgs} by analyzing the strongly coupled regime of conformal field theories in $1/N$ expansion. This becomes possible because of recent developments on conformal bootstrap technique, e.g., in \cite{ElShowk:2012ht}.}
where 2d W$_N$ minimal model  is dual to  Prokushkin-Vasiliev theory on AdS$_3$ given by \cite{Prokushkin:1998bq}. We examine three point functions with two scalar operators and one higher spin current in the minimal model up to the next leading order in $1/N$ expansion. They should be interpreted as one-loop corrections to three point interactions between two bulk scalars and one higher spin gauge fields in the dual higher spin theory. 
We develop a simple and systematic method to compute the three point functions by decomposing four point functions of scalar operators with Virasoro conformal blocks. 
Among others, we expect that this way of computation makes  the dual higher spin interpretation easier.
Applying the method, we reproduce known results at the leading order in $1/N$ obtained by
 \cite{Chang:2011mz,Ammon:2011ua}. Exact results are available only up to correlators with spin $5$ current \cite{Bais:1987zk,Ahn:2011by,Ahn:2013sua}, but a simple relation was conjectured for generic $s$ in \cite{Ahn:2013sua}. We obtain the $1/N$ corrections of correlators with spin $s \leq 8$ current, and the results for $s =6,7,8$ should be new. We check that they satisfy the conjectured relation as confirmation of our results.

We would like to examine the W$_N$ minimal model in $1/N$ expansion, 
but we should specify the expansion in more details.
The minimal model has a coset description
\begin{align}
	\frac{\text{su}(N)_k \oplus \text{su}(N)_1}{\text{su}(N)_{k+1}} \, ,
	\label{coset}
\end{align}
whose central charge is given by
\begin{align}
c &= (N-1) \left( 1 - \frac{N (N+1)}{(N+k) (N+k+1)}\right) \, .
\label{center}
\end{align}
The model has two parameters $N,k$.
For our purpose, it is convenient to define the 't~Hooft coupling
\begin{align}
\lambda = \frac{N}{N+k}
\label{thooft}
\end{align}
and label the model by $N,\lambda$ instead of $N,k$.
We then expand the model in $1/N$, where each order depends on the other parameter $\lambda$.  The expansion is almost the same as $1/c$ expansion because of $c \sim N (1 - \lambda^2) + \mathcal{O}(N^0)$, but details are different.

The minimal model is argued to be dual to the higher spin theory of \cite{Prokushkin:1998bq}, which includes higher spin gauge fields $\varphi^{(s)}$ $(s=2,3,4,\ldots)$ and complex scalar fields $\phi_\pm$ with mass $m^2 = - 1 + \lambda^2$.
The large $N$ limit of minimal model with $\lambda$ in \eqref{thooft} kept finite corresponds to the classical limit of higher spin theory, where $\lambda$ is identified with the parameter in bulk scalar mass. The higher spin gauge fields $\varphi^{(s)}$ and bulk scalars $\phi_\pm$ are dual to higher spin currents $J^{(s)}$ and scalar operators $\mathcal{O}_\pm$, respectively. Here different boundary conditions are assigned to the bulk scalars $\phi_\pm$ and the dual conformal dimensions are given by $\Delta_\pm = 2 h_\pm = 1 \pm \lambda$ at the tree level.

Basic data of conformal field theory may be given by spectrum and three point functions of primary operators. Since higher spin symmetry of the minimal model is exact, spectrum does not receive any corrections in $1/N$. Namely, there is no anomalous dimension for higher spin current $J^{(s)}$. Therefore, as simple but non-trivial examples, we examine three point functions and specifically focus on those with two scalar operators and one higher spin current
as 
\begin{align}
\langle \mathcal{O}_\pm (z_1) \bar{\mathcal{O}}_\pm (z_2) J^{(s)} (z_3) \rangle 
\label{3pt0}
\end{align}
with $s=2,3,4,\ldots$.
Here $\bar{\mathcal{O}}_\pm$ are complex conjugate of $\mathcal{O}_\pm$.
In \cite{Chang:2011mz,Ammon:2011ua}, the three point functions in the large $N$ limit of the minimal model have been computed from classical higher spin theory.
They were reproduced with conformal field theory approach in \cite{Ammon:2011ua,Creutzig:2012xb,Moradi:2012xd},%
\footnote{The analysis in \cite{Creutzig:2012xb,Moradi:2012xd} were made in the context of $\mathcal{N}=2$ holographic duality in \cite{Creutzig:2011fe}, but we can see that the analysis reduces to that for the bosonic case under a suitable truncation at the leading order in $1/N$.}
but these methods are applicable only to the leading order analysis in $1/N$.
Since the W$_N$ minimal model  is solvable, for instance, by making use of the coset description \eqref{coset}, we can obtain the three point functions \eqref{3pt0} with finite $N,k$ in principle. However, in practice, the computation would be quite complicated, and only explicit expressions are available only with spin $3,4,5$ currents \cite{Bais:1987zk,Ahn:2011by,Ahn:2013sua} (see also \cite{Gaberdiel:2012ku} for an alternative algebraic method).

In this paper, we develop a different way to compute the three point functions \eqref{3pt0} from the  decomposition of scalar four point functions by Virasoro conformal blocks.
Our method may be explained as follows;
Let us consider a generic operator product expansion of scalar operators $\mathcal{O}_{i}$ with conformal weights $(h_i,h_i)$ as
\begin{align}
	\mathcal{O}_1 (z_1) \mathcal{O}_2 (z_2)  = \sum_p   \frac{C_{12 p}}{z_{12}^{h_1 + h_2 - h_p} {\bar z}_{12}^{h_1 + h_2 - \bar{h}_p} } \mathcal{A}_p (z_2) + \cdots \, ,
	\label{AbsOPE}
\end{align}
where the coefficient $C_{12 p}$ includes the information of three point function.
Moreover, $\mathcal{A}_p$ has conformal weights $(h_p , \bar{h}_p)$, and dots denote contributions from descendants. Using the expansion, we can decompose scalar four point function as
\begin{align}
\langle \mathcal{O}_{1} (\infty) \mathcal{O}_{2} (1) \mathcal{O}_{3} (z) \mathcal{O}_{4} (0) \rangle = \sum_p  \frac{ C_{12p} C_{34 p}}{|z|^{2 (h_3 + h_4) }  } \mathcal{F} (c,h_i ,h_p , z) \bar{ \mathcal{F}} (c,h_i ,\bar h_p , \bar z ) \, .
\label{AbsCPWE}
\end{align}
Here $\mathcal{F} (c,h_i , h_p, z) $ is Virasoro conformal block, which can be fixed only from the symmetry in principle. Once we know scalar four point functions and Virasoro conformal blocks, we can read off  coefficients as $C_{12 p}$ by solving constraint equations coming from \eqref{AbsCPWE}.
For our case with $\mathcal{O}_i = \mathcal{O}_\pm$ or $\bar{\mathcal{O}}_\pm$,
four point functions can be computed exactly with finite $N, k$, for instance, by applying Coulomb gas approach as in \cite{Papadodimas:2011pf}. 
 On the other hand,  Virasoro conformal blocks are quite complicated, but explicit forms may be obtained by applying Zamolodchikov's recursion relation \cite{Zamolodchikov:1985ie},
 see also \cite{Perlmutter:2015iya,Beccaria:2015shq}.
We can find other works on the $1/c$ expansion of Virasoro conformal blocks in, e.g., \cite{Fitzpatrick:2015zha,Fitzpatrick:2015dlt,Chen:2016cms,Fitzpatrick:2016mtp}.
Gathering these knowledges, we shall obtain the coefficients as $C_{12 p}$ up to the next leading order in $1/N$ expansion.

The paper is organized as follows;
In order to study the decomposition \eqref{AbsCPWE}, we need to examine scalar four point functions and Virasoro conformal blocks. In the next section we  decompose scalar four point functions in terms of cross ratio $z$, and in section  \ref{VB} we give the explicit expressions of Virasoro conformal blocks in  expansions both in $1/N$ and $z$.
After these preparations, we compute three point functions \eqref{3pt0} by solving constraint equations coming from \eqref{AbsCPWE} in section \ref{3ptfn}.
In subsection \ref{tree} we reproduce known results at the leading order in $1/N$.
In subsection \ref{1/Ncorrections} we obtain the $1/N$ corrections of three point functions for $s=3,4,\ldots,8$, and check that they satisfy the relation conjectured in \cite{Ahn:2013sua}.  
Section \ref{conclusion} is devoted to conclusion and discussions.
In appendix \ref{Zamolodchikov} we examine Virasoro conformal blocks in expansions of $1/c$ and $z$ by analyzing Zamolodchikov's recursion relation. In appendix \ref{double} we compute three point functions with higher spin currents of double trace type.

\section{Expansions of four point functions}
\label{Expansions}

We would like to obtain the coefficients as $C_{12p}$ by solving \eqref{AbsCPWE}. 
For the purpose, we need information on the both sides of the equation, i.e., scalar four point functions and Virasoro conformal blocks. In this section we examine scalar four point functions.
We are interested in three point functions of two scalar operators $\mathcal{O}_\pm$ and a higher spin current $J^{(s)}$ as in \eqref{3pt0}.
We consider the following four point functions with scalar operators $\mathcal{O}_\pm$ as
\begin{align}
G_{++} (z) \equiv \langle \mathcal{O}_+ (\infty) \bar{\mathcal{O}}_+ (1) \mathcal{O}_+ (z) \bar{\mathcal{O}}_+ (0) \rangle \label{G++0}\, , \\
G_{--} (z) \equiv \langle \mathcal{O}_- (\infty) \bar{\mathcal{O}}_- (1) \mathcal{O}_- (z) \bar{\mathcal{O}}_- (0) \rangle \label{G--0}\, , \\
G_{-+} (z) \equiv \langle \mathcal{O}_- (\infty) {\mathcal{O}}_+ (1) \bar{\mathcal{O}}_+ (z) \bar{\mathcal{O}}_- (0) \rangle \, . \label{G-+0} 
\end{align}
Exact expressions with finite $N,k$ may be found in \cite{Papadodimas:2011pf}.
From the expansions in $z$, we can read off what kind of operators are involved in the decomposition by Virasoro conformal blocks. In the rest of this section, we obtain the explicit forms of four point functions in $z$ expansion for  parts relevant to later analysis.

Let us first examine the $z$ expansion of $G_{++}(z)$ in \eqref{G++0},
and see generic properties of the four point functions.
The expression with finite $N,k$ is \cite{Papadodimas:2011pf} 
\begin{align}
G_{++} (z)  = |z (1-z)|^{-2 \Delta_+} \left[ \left| (1 - z)^{1 + \lambda} 
{}_2 F_1 \left(1 + \frac{\lambda}{N} , - \frac{\lambda}{N} ; - \lambda ; z \right)  \right|^2
\right.  \nonumber
\\ \left.  + \mathcal{N}_1  \left| z^{1 + \lambda} 
{}_2 F_1 \left(1 + \frac{\lambda}{N} , - \frac{\lambda}{N} ; 2 + \lambda ; z \right)  \right|^2 \right] 
\label{G++}
\end{align}
with
\begin{align}
\mathcal{N}_1 = - \frac{\Gamma(1 + \lambda - \frac{\lambda}{N}) \Gamma(- \lambda) ^2 \Gamma(2 + \lambda + \frac{\lambda}{N})}{\Gamma(- 1 - \lambda - \frac{\lambda}{N}) \Gamma(\frac{\lambda}{N} - \lambda) \Gamma(2 + \lambda)^2} \, .
\end{align}
Here the exact value of conformal dimensions $\Delta_+ = 2 h_+$ is 
\begin{align}
\Delta_+ = \frac{(N-1)(2 N+1+ k)}{N (N+k)} = 1 + \lambda - \frac{1}{N} - \frac{1}{N^2} \lambda  + \mathcal{O} (N^{-3})\, ,
\label{Delta+}
\end{align}
which is expanded in $1/N$ up to the $N^{-2}$ order.

In the expansion in $z$, we would like to pick up the terms corresponding to the three point function \eqref{3pt0}. 
The operator product of $\mathcal{O}_+$ may be expanded as
\begin{align}
 \label{OPEO}
\mathcal{O}_+ (z) \bar{\mathcal{O}}_+(0)
= & \frac{1}{|z|^{2 \Delta_+ }} 
 + \sum_s  \frac{C_{+}^{(s)} z^s}{|z|^{2 \Delta_+ } }   J^{(s)} (0)
 \\ & + \sum_{(s_1,s_2;s')}  \frac{C_{+}^{(s_1,s_2;s' )} z^{s' }}{|z|^{2 \Delta_+ } }   J^{(s_1 ,s_2 ; s')} (0)
 + \sum_{n,m} C_{+}^{(n,m)}   z^n \bar z^m \mathcal{A}_{(n,m)} (0)
 \cdots \, . \nonumber 
\end{align}
Here $J^{(s_1,s_2;s')} (z)$ are higher spin currents of double trace type as
\begin{align}
 J^{(s_1 ,s_2 ; s')} = J^{(s_1)} \partial^{s ' - s_1 - s_2} J^{(s_2)} + \cdots 
\end{align}
with $ s' \geq 6$ as $s_1, s_2 \geq 3$ and $s' - s_1 - s_2  \geq 0$.
If we use the normalization as $\langle J^{(s)} J^{(s)} \rangle \propto N$, then the two point function of this type of operator becomes $\langle J^{(s_1,s_2;s')}J^{(s_1,s_2;s')} \rangle \propto N^2$.
This is related to the fact that $C_+^{(s)} \propto N^{-1/2}$, while 
$C_+^{(s_1,s_2;s)} \propto N^{-1}$. There could be currents of other multi-trace type, but the contributions are more suppressed in $1/N$.
Furthermore, $\mathcal{A}_{(n,m)}(z)$ are double trace type operators of the form as
\begin{align}
\mathcal{A}_{(n,m)} = \mathcal{O}_+ \partial^n \bar{\partial}^m \bar{\mathcal{O}}_+ +  \cdots \, ,
\end{align}
and  the conformal weights  are $(h_{n,m} , \bar h_{n,m}) = (2 h_+ + n , 2 h_+ + m)$.
The dots in \eqref{OPEO} include the operators dressed by higher spin currents $J^{(s)} (z), \bar J^{(s)} (\bar z)$ for instance.

The operator product expansion in \eqref{OPEO} suggests that the contributions from $J^{(s)}$ or its descendants are included in terms like $z^{s + l } / |z|^{  2 \Delta_+}$, where $l = 0,1,2, \ldots$ corresponds to the level of descendant. In \eqref{G++}, such terms appear as
\begin{align}
G_{++} (z)  = |z|^{-2 \Delta_+}  (1 - z)^{ - \Delta_+ + 1 + \lambda} 
{}_2 F_1 \left(1 + \frac{\lambda}{N} , - \frac{\lambda}{N} ; - \lambda ; z \right)  + \cdots \, .
\label{rG++}
\end{align}
Note that they also  include effects from higher spin currents of double trace type  $J^{(s_1 ,s_2 ; s')} (z)$ among others.
For the first term in \eqref{G++}, the other contributions involve at least one anti-holomorphic current ${\bar J}^{(s)} (\bar z)$. For the second term  in \eqref{G++}, the expansions become polynomials of $z$ and $\bar z$ at the leading order in $1/N$, and this implies that double trace type operators $\mathcal{A}_{(n,m)}$ should appear as $\mathcal{A}_p$ in \eqref{AbsOPE}.
At the leading order in $1/N$, we can expand  \eqref{rG++} around $z \sim 0$ as
\begin{align}
G_{++} (z) \sim |z|^{-2 (\lambda + 1)}   \, .
\end{align}
This corresponds to the expansion by the identity operator in  \eqref{OPEO}.
Thus the non-trivial contributions to our three point functions come form the terms at least of order $1/N$.

At the next and next-to-next orders in $1/N$, there are two types of contributions in \eqref{rG++}.
One  comes from
\begin{align}
(1-z)^{ -  \Delta_+ + 1 + \lambda} = (1-z)^{\frac{1}{N} +  \frac{\lambda}{N^2}} + \mathcal{O}(N^{-3}) \, ,
\end{align}
which becomes
\begin{align}
&(1-z)^{\frac{1}{N}} (1-z)^{\frac{\lambda}{N^2}} \nonumber \\
& \qquad = 1 - \frac{1}{N} \sum_{k=1}  \frac{1}{k}z^k
+ \frac{1}{N^2} \left[  \sum_{k=1}^\infty \left( - \frac{\lambda}{k} z^k \right)
+ \sum_{k=2}^\infty \frac{1}{k} H_{k-1} z^k \right] + \mathcal{O}(N^{-3}) \, .
\end{align} 
Here we have used for $k \geq 2$
\begin{align}
\binom{\frac{1}{N}}{k} &=
\frac{\Gamma(1+ \frac{1}{N})}{k! \Gamma(1-k + \frac{1}{N})}
= (-1)^{k-1} \frac{1}{N k!} \left( 1 - \frac{1}{N}\right) \left( 2 - \frac{1}{N}\right) \cdots  
\left( k-1 - \frac{1}{N}\right) \nonumber \\
&= (-1)^{k-1} \frac{1}{Nk} \left( 1 - \frac{1}{N} \sum_{i=1}^{k-1} \frac{1}{i} \right)  + \mathcal{O}(N^{-3}) \, ,
\end{align}
and the definition of harmonic number
\begin{align}
H_n = \sum_{j=1}^n \frac{1}{j} \, .
\end{align}
The other comes from the hypergeometric function,
which can be similarly expanded in $1/N$  as
\begin{align}
&{}_2 F_1 \left(1 + \frac{\lambda}{N} , - \frac{\lambda}{N} , - \lambda ; z \right)
= \frac{\Gamma (- \lambda)}{\Gamma(1 + \tfrac{\lambda}{N}) \Gamma (-\frac{\lambda}{N})}
\sum_{n=0}^\infty \frac{\Gamma(1 + \frac{\lambda}{N} + n) \Gamma(- \frac{\lambda}{N} + n)}{\Gamma(- \lambda + n)} \frac{z^n}{n!}  \\
&=1 + \frac{\Gamma(1 - \lambda)}{N}  \sum_{n=1}^\infty \frac{\Gamma (n)}{\Gamma(n-\lambda)} z^n   + \frac{1}{N^2} \left(  \lambda z + \lambda \Gamma(1 - \lambda) \sum_{n=2}^\infty \frac{\Gamma (n)}{n \Gamma(n-\lambda)} z^n 
\right)
+ \mathcal{O}(N^{-3})  \, . \nonumber
\end{align}
In total, we have
\begin{align}
|z|^{ 2 \Delta_+} G_{++} (z) \sim   1 + \frac{1}{N} \sum_{n=1}^\infty \left( - \frac{1}{n} +   \frac{ \Gamma(1 - \lambda) \Gamma (n)}{\Gamma(n-\lambda)}  \right)z^n  
+ \frac{1}{N^2} \sum_{n=2}^\infty  f^{(n)}_{++} z^n \, ,
\label{ExpG++}
\end{align}
where
\begin{align}
\label{fn++}
f^{(n)}_{++} =  - \sum_{l=1}^{n-1} \frac{\Gamma(1 - \lambda) \Gamma(l)}{(n-l)\Gamma(l - \lambda)} - \frac{\lambda}{n} + \frac{H_{n-1}}{n} 
+ \frac{\lambda \Gamma(1-\lambda) \Gamma(n)}{n \Gamma(n-\lambda)} \, . 
\end{align}
First few expressions are
\begin{align}
f^{(2)}_{++} &=\frac{1}{2} \left( -2 - \frac{1}{\lambda - 1} - \lambda \right) \, , \label{f2++}\\
f^{(3)}_{++} &= \frac{1}{3} \left( \frac{4}{\lambda - 2} + \frac{1}{\lambda - 1} - \lambda \right)
\, ,\\
f^{(4)}_{++}&= \frac{1}{8} \left( 1 - \frac{18}{\lambda - 3} + \frac{8}{\lambda -2} + \frac{14}{\lambda - 1}  - 2 \lambda \right) \, .
\end{align}

We would like to move to another four point function $G_{--}(z)$ in \eqref{G--0},
whose expression with finite $N,k$ can be again found in \cite{Papadodimas:2011pf}. 
We use the four point function in order to obtain the three point function \eqref{3pt0} with 
the other type of scalar operator $\mathcal{O}_-$.
As for $G_{++}(z)$, the relevant part is 
\begin{align}
G_{--} (z) = |z|^{-2 \Delta_- } (1-z)^{\frac{1}{N} -\frac{\lambda}{N^2}}
{}_2 F_1 \left( 1 - \frac{\lambda}{N}, \frac{\lambda}{N} - \frac{\lambda^2}{N^2} ; \lambda - \frac{\lambda^2}{N} ; z \right) + \cdots \, .
\end{align}
Here we may need
\begin{align}
\Delta_- &= 2 h_- = \frac{N-1}{N} \left( 1 - \frac{N+1}{N+k+1} \right)  \nonumber \\
 &= 1 - \lambda - \frac{1}{N} (1 - \lambda ^2) + \frac{1}{N^2} \lambda( 1 - \lambda^2) + \mathcal{O} (N^{-3}) \, .
 \label{Delta-}
\end{align}
Similarly to $G_{++}(z)$ we can expand $G_{--} (z)$ in $z$ as
\begin{align}
|z|^{ 2 \Delta_- } G_{--} (z) \sim   1 + \frac{1}{N} \sum_{n=1}^\infty  \left( - \frac{1}{n} +   \frac{ \Gamma(1 + \lambda) \Gamma (n)}{\Gamma(n + \lambda)}  \right) z^n 
+ \frac{1}{N^2} \sum_{n=2}^\infty  f^{(n)}_{--} z^n \, ,
\label{ExpG--}
\end{align}
where
\begin{align}
\label{fn--}
f^{(n)}_{--} =  - \sum_{l=1}^{n-1} \frac{\Gamma(1 + \lambda) \Gamma(l)}{(n-l)\Gamma(l + \lambda)} + \frac{\lambda}{n} + \frac{H_{n-1}}{n} 
+ \frac{\lambda \Gamma(1+\lambda) \Gamma(n)}{\Gamma(n+\lambda)} \left( \sum_{k=1}^{n-1} \frac{\lambda}{k + \lambda}  - \frac{1}{n} \right)\, . 
\end{align}
First few expressions are
\begin{align}
f^{(2)}_{--} &=\frac{\lambda}{2}-\frac{3}{2 (\lambda+1)}+\frac{1}{(\lambda+1)^2} \, , \label{f2--}\\
f^{(3)}_{--} &=\frac{\lambda}{3}-\frac{13}{3 (\lambda+1)}+\frac{2}{(\lambda+1)^2}+\frac{20}{3 (\lambda+2)}-\frac{8}{(\lambda+2)^2}
\, ,\\
f^{(4)}_{--}&= \frac{\lambda}{4}-\frac{31}{4 (\lambda+1)}+\frac{3}{(\lambda+1)^2}+\frac{23}{\lambda+2}-\frac{24}{(\lambda+2)^2}-\frac{63}{4 (\lambda+3)}+\frac{27}{(\lambda+3)^2}+\frac{1}{8}\, .
\end{align}

From the four point functions $G_{\pm \pm} (z)$, we can read off the square root of  coefficients $(C_{\pm}^{(s)})^2$, but relative phase factor cannot be fixed.
In order to determine it, we also need to examine $G_{-+} (z)$ in \eqref{G-+0}, which 
can be computed as \cite{Papadodimas:2011pf}
\begin{align}
G_{-+}(z) =|1-z|^{-2\Delta_+}|z|^{\frac{2}{N}}\left|1+\frac{1-z}{Nz}\right|^2
\end{align}
with finite $N,k$. For later arguments, we need 
\begin{align}
|1-z|^{2\Delta_+} G_{-+}(z) \sim 1+\frac{1}{N}\sum^\infty_{n=2}\frac{n-1}{n}(1-z)^n \, ,
\label{ExpG-+}
\end{align}
which is expanded in $(1-z)$ up to the $1/N$ order.

\section{Virasoro conformal blocks}
\label{VB}

In the previous section we analyzed the left hand side of \eqref{AbsCPWE}. In order to obtain three point functions by solving the equations in \eqref{AbsCPWE}, we further need information on  $\mathcal{F}(c,h_i , h_p , z)$.
In general, the forms of Virasoro conformal blocks are quite complicated.
In practice, we actually do not need to know closed forms but expansions in $z$ up to some orders.
For the purpose, a standard approach may be  solving Zamolodchikov's recursion relation in \cite{Zamolodchikov:1985ie}. Following the algorithm developed in \cite{Perlmutter:2015iya} (see also \cite{Beccaria:2015shq}), we obtain the expressions of Virasoro conformal blocks to several orders in $z$ and $1/c$ in appendix \ref{Zamolodchikov}.
Related works may be found in \cite{Fitzpatrick:2015zha,Beccaria:2015shq,Fitzpatrick:2015dlt,Chen:2016cms,Fitzpatrick:2016mtp},
and in particular, some closed form expressions were given, e.g., in \cite{Fitzpatrick:2015zha}.
Our findings agree with their results after minor modifications

Le us consider the four point function in the decomposition of \eqref{AbsCPWE} with $h_1 = h_2$ and $h_3 = h_4$.
In the decomposition, intermediate operator $\mathcal{A}_p$ can be the identity or other. As observed in the examples of previous section, only the Virasoro conformal block with the identity operator (called as vacuum block) survives at the leading order in $1/N$. This simply means that the four point functions are factorized into the products of two point ones at the leading order in $1/N$. Virasoro conformal block with $\mathcal{A}_p$ of single trace type would appear at the next leading order in $1/N$. We would like to examine $1/N$ corrections to three point functions, so we  need $1/N$ corrections to the Virasoro block of  $\mathcal{A}_p$.
This also implies that we need the expression of vacuum block up to the next-to-next leading order in $1/N$.

Let us first examine the vacuum block with $h_1 = h_3 = h_\pm$.
As was explained in appendix \ref{Zamolodchikov}, the $1/c$-expansion of vacuum block is given by
\begin{align}
\label{vbexp}
\mathcal{V}_0 (x)  = &1 + \frac{2 h_1 h_3 }{c} z^2 {}_2 F_1 (2,2;4;z)  \\
&  + \frac{1}{c^2} \left [ h_1^2 h_3^2 k_a (z)   + h_1 h_3 (h_1 + h_3)  k_b (z)+ h_1 h_3 k_c (z) \right ]+ \mathcal{O} (c^{-3})  \nonumber 
\end{align}
with
\begin{align}
& k_a (z)  = 2 z^4  +4 z^5  +\frac{28 z^6}{5}+\frac{34 z^7}{5}  +\frac{2687 z^8}{350} + \mathcal{O}(z^9) \, , \nonumber \\
&k_b (z) = \frac{2 z^4}{5} +\frac{4 z^5}{5}+\frac{39 z^6}{35}+\frac{47 z^7}{35}+\frac{263 z^8}{175}+ \mathcal{O}(z^9) \, , \label{kb} \\
&k_c (z) = \frac{2 z^4}{25} +\frac{4 z^5}{25} +\frac{109 z^6}{490} +\frac{131 z^7}{490}+ \frac{1879 z^8}{6300} +  \mathcal{O}(z^9) \, . \nonumber
\end{align}
The $1/c$ order term  corresponds to the exchange of spin 2 current (energy momentum tensor) in terms of global block. 
We need to rewrite the expansion in $1/c$ by that in $1/N$ as
\begin{align}
\label{Vacexp}
\mathcal{V}_0 (z) =   \mathcal{V}_0^{(0)} (z) +  \mathcal{V}_0^{(1)} (z) \frac{1}{N} + \mathcal{V}_0^{(2)} (z) \frac{1}{N^2} + \mathcal{O} (N^{-3}) \, .
\end{align}
The first two terms can be easily read off as
\begin{align}
\label{V0}
\mathcal{V}_0^{(0)} (z)= 1 \, , \quad
\mathcal{V}_0^{(1)} (z)= \frac{1}{2} \left( \frac{1 \pm \lambda}{1 \mp \lambda} \right)  z^2 
{}_2 F_1 (2,2;4;z)\, . 
\end{align}
Since there are two types of contributions to $\mathcal{V}_0^{(2)}$, we separate it into two parts as 
\begin{align}
\mathcal{V}_0^{(2)} = \mathcal{V}^{(2,1)}_0 (z) + \mathcal{V}^{(2,2)}_0 (z) \, .
\label{vacuumN2}
\end{align}
One comes from the $1/c$ order term  with the next leading contribution from $h_\pm^2/c$ as
\begin{align}
\nonumber
\mathcal{V}^{(2,1)}_0 (z) &= 
f_{\pm \pm}^{(2) }  z^2 {}_2 F_1 (2,2;4;z)  \\
&=f_{\pm \pm}^{(2) }  
\left( z^2 + z^3 + \frac{9z^4}{10} + \frac{4z^5}{5} + \frac{5 z^6}{7} + \frac{9 z^7}{14} + \frac{7 z^8}{12} 
\right) + \mathcal{O} (z^{9}) \, ,
\label{vacuumN21}
\end{align}
where $f_{\pm \pm}^{(2)}$ were given in \eqref{f2++} and \eqref{f2--}.
Here we have used 
\begin{align}
&\frac{2 h_\pm^2}{c} = \frac{1}{2N} \left( \frac{1 \pm \lambda}{1 \mp \lambda} \right) 
+ \frac{1}{ N^2} f_{\pm \pm}^{(2)} + \mathcal{O} (N^{-3}) \, ,
\label{cwexp}
\end{align}
which are obtained from the $1/N$ expansions of $h_\pm$ as in \eqref{Delta+} and \eqref{Delta-} and $c$ in \eqref{center}  as
\begin{align}
&c 
= N (1- \lambda^2) \left[1 - \frac{1}{N} \left(\lambda + \frac{1}{1+\lambda}\right) \right] + \mathcal{O} (N^{-1})\, .
\end{align}
The other comes from the $1/c^2$ order terms in \eqref{vbexp} as
\begin{align}
 \mathcal{V}_0^{(2,2)} (z)= \frac{(1 \pm \lambda)^2}{16(1 \mp \lambda)^2} k_a (z) 
+ \frac{(1 \pm \lambda)}{4(1 \mp \lambda)^2} k_b (z)
  + \frac{1}{4 (1 \mp \lambda)^2} k_c (z)
\label{vacuumN22p}
\end{align}
with $k_a (z)$, $k_b (z)$, and $k_c (z)$ in \eqref{kb}.

We also need Virasoro blocks of $\mathcal{A}_p$ up to the next non-trivial order in $1/N$. 
It is known that the Virasoro block is expanded in $1/c$ as (see, e.g., \cite{Fitzpatrick:2016mtp})
\begin{align}
\mathcal{V}_p (z)= g(h_p ,z) + \frac{1}{c} \left[ h_1 h_3 f_a (h_p,z) + (h_1 + h_3) f_b (h_p,z) +  f_c (h_p,z)  \right] + \mathcal{O} (c^{-2}) \, .
\label{vbop0}
\end{align}
Here $g(h_p,z)$ is the global block of $\mathcal{A}_p$ and the expressions of $f_a (h_p,z)$, $f_b (h_p,z)$, and $f_c (h_p,z)$ were obtained in 
\cite{Fitzpatrick:2016mtp}. See also appendix \ref{Zamolodchikov}. 
For our application, we set $h_1 = h_3 = h_\pm$ and $h_p = s$. 
We need the expansion in $1/N$ instead of $1/c$ as
\begin{align}
\label{vbop}
\mathcal{V}_s (z) =   \mathcal{V}_s^{(0)} (z) +  \mathcal{V}_s^{(1)} (z) \frac{1}{N}  + \mathcal{O} (N^{-2}) \, .
\end{align}
The leading term $\mathcal{V}_p^{(0)} (z) $ is given by the global block as
\begin{align}
\mathcal{V}_s^{(0)} (z) = g(s , z) = z^s {}_2 F_1 (s,s;2s ; x) \, .
\label{vbop1}
\end{align}
The next order contributions in $1/N$ are 
\begin{align}
\mathcal{V}_s^{(1)} (z) = \frac{1}{4 } \frac{1 \pm \lambda}{1\mp \lambda} f_a (s , z)  + 
\frac{1}{(1 \mp \lambda)} f_b (s ,z) + \frac{1}{(1 - \lambda^2)} f_c (s,z) \, ,
\label{vbop2}
\end{align}
where the functions $f_a (s , z)$, $f_b (s , z)$, and $ f_c (s , z) $ are given by
\begin{align}
f_a (s , z) 
&  =z^s \left[ 2 z^2 +(s+2) z^3 +\frac{(s+3) (5 s (s+3)+6) z^4}{20 s+10} + \frac{(s+3) (s+4) (5 s (s+4)+8) z^5}{60 (2 s+1)} \right. \nonumber \\ & \left.    + \mathcal{O}(z^{6})    \right] \nonumber \, , \\
f_b (s ,z) 
&  =  s (s-1) z^s  \left[\frac{z^2}{2 s+1} +\frac{(s+2) z^3}{4 s+2} +\frac{(s+3) (5 s (s+4)+18) z^4}{20 (4 s (s+2)+3)} \right.  \nonumber  \\ & \left. +  \frac{(s+3) (s+4) (5 s (s+5)+24) z^5}{120 (4 s (s+2)+3)} + \mathcal{O}(z^6)  \right]  \, , \label{fb} \\
f_c (s,z) 
&=  \frac{s^2 (s-1)^2}{2 (2s+1)^2} z^s 
\left[ z^2 + \frac{(s+2)z^3 }{2}+ \frac{(s+3) (s (10s (2s+11) + 191) + 108)z ^4}{40 (2s+3)^2} \right.
\nonumber \\
&+ \left. \frac{(s+3)(s+4)(s(10s (2s+13)+243)+144)z^5}{240(2s+3)^2} + \mathcal{O}(z^6) 
\right] \, . \nonumber 
\end{align}

\section{Three point functions}
\label{3ptfn}

After the preparations in previous sections, we  now work on the decompositions of four point functions by Virasoro conformal blocks as in \eqref{AbsCPWE}.
In the current case, the decompositions are
\begin{align}
|z |^{2 \Delta_\pm} G_{\pm \pm }(z) =  \mathcal{V}_0 (z) + \sum_{s=3}^\infty (C_\pm^{(s)})^2 \mathcal{V}_s (z) + \sum_{(s_1,s_2;s')} (C_\pm^{(s_1 ,s_2 ;s')})^2 \mathcal{V}_{ s'} (z)   + \cdots \, .
\label{highercond}
\end{align}
Here $G_{\pm \pm} (z)$ are four point functions defined in \eqref{G++0} and \eqref{G--0}, and the expansions in $z$ were obtained as in \eqref{ExpG++} and \eqref{ExpG--}.
Moreover, $\mathcal{V}_0 (z) $ is the vacuum block and $\mathcal{V}_s (z) $ is the Virasoro block of higher spin current $J^{(s)}$ (or $J^{(s_1,s_2;s)}$). Their expansions in $z$ can be found in the previous section.

Solving constraint equations from \eqref{highercond}, we  read off 
the coefficients $C_\pm^{(s)}$, which are proportional to the  three point functions \eqref{3pt0}.
It is convenient to expand the coefficients in $1/N$ as
\begin{align}
C_\pm^{(s)} = \frac{1}{N^{1/2}} \left( C_{\pm ,0}^{(s)} + \frac{1}{N} C_{\pm ,1}^{(s)} + \mathcal{O} (N^{-2}) \right) \, .
\end{align}
Then we can see that the constraint equations from \eqref{highercond} at the order $N^0$ is trivially satisfied as $1=1$.
The non-trivial conditions arise from order $1/N$ terms, and they determine the leading order expressions $C_{\pm ,0}^{(s)}$ as seen in the next subsection.
The main purpose of this paper is to compute $C_{\pm ,1}^{(s)}$, which are $1/N$ corrections to the leading order expressions. We derive them by solving order $N^{-2}$ conditions up to $s=8$ in subsection \ref{1/Ncorrections}. Notice that we should take care of $C_\pm^{(s_1 ,s_2 ; s')}$ in \eqref{highercond} for  $s \geq 6$,  which may be expanded as
\begin{align}
C_\pm^{(s_1 , s_2 ; s')} = \frac{1}{N} C^{(s_1,s_2;s')}_{\pm ,0}  + \mathcal{O} (N^{-2}) \, .
\end{align}
The coefficients $C^{(s_1,s_2;s')}_{\pm ,0} $ are analyzed in appendix \ref{double}.

\subsection{Leading order expressions in $1/N$}
\label{tree}

We start from three point functions at the leading order in $1/N$.
We examine the constraint equations from \eqref{highercond} up to $1/N$ order.
Up to this order, the vacuum block is given by (see \eqref{vbexp})
\begin{align}
\mathcal{V}_0 (z) =  1 + \frac{1}{N} (C_{\pm,0}^{(2)})^2 z^2 {}_2 F_1 (2,2;4;z) + \mathcal{O} (N^{-2}) \, ,
\end{align}
where we have defined
\begin{align}
\label{C3f}
C_{\pm,0}^{(2)}  =  \left. \sqrt{\frac{ 2 (h_\pm )^2}{c}} \right|_{\mathcal{O}(N^{-1/2})}=   \sqrt{\frac{1}{2} \frac{1 \pm \lambda}{1 \mp \lambda}} \, .
\end{align}
The Virasoro block of $J^{(s)}$ is 
\begin{align}
\mathcal{V}_s (z) = z^s {}_2 F_1 (s,s;2s ; z) + \mathcal{O} (N^{-1}) 
\end{align}
as in \eqref{vbop} with \eqref{vbop1}.
Therefore, the expansion in  \eqref{highercond} can be written as
\begin{align}
|z|^{ 2 \Delta_\pm} G_{\pm \pm } (z) = 
  1 + \frac{1}{N} \sum_{s=2}^\infty (C_{\pm,0}^{(s)})^2 z^s {}_2 F_1 (s,s;2s ; z) + \cdots 
\label{CPWE1}
\end{align}
up to the order of $1/N$.
The four point functions $G_{\pm \pm} (z)$ can be expanded  as
\begin{align}
|z|^{ 2 \Delta_\pm } G_{\pm\pm} (z) \sim   1 + \frac{1}{N} \sum_{n=1}^\infty z^n  \left( - \frac{1}{n} +   \frac{ \Gamma(1 \mp \lambda) \Gamma (n)}{\Gamma(n \mp \lambda)}  \right)
+ \cdots 
\end{align}
as in \eqref{ExpG++} and \eqref{ExpG--} up to the same order.
On the other hand, the global blocks can be written as
\begin{align}
 z^s {}_2 F_1 (s,s;2s;z) 
=  \frac{\Gamma(2s)}{(\Gamma(s))^2} \sum_{n=0}^\infty \frac{(\Gamma(s+n))^2}{\Gamma(2s + n)} \frac{z^{n + s}}{n!} \, .
\end{align}
Comparing the coefficients in front of $z^n$, we obtain 
\begin{align}
 - \frac{1}{n} + \frac{\Gamma(1 \mp \lambda) \Gamma(n)}{\Gamma(n \mp \lambda)}
= (\Gamma(n))^2 \sum_{s=2}^n \frac{\Gamma(2s) (C_{\pm,0}^{(s)})^2}{(\Gamma(s))^2 \Gamma(s+n) (n-s)!} \, .
\label{consistent}
\end{align}
They are the constraint equations for $(C_{\pm,0}^{(s)})^2$ with  $s =3,4,\ldots$.

In order to fix  relative phase factor, we examine $G_{-+}(z)$ in \eqref{G-+0} as well.
The decomposition in \eqref{AbsCPWE} become
\begin{align}
|1-z|^{2\Delta_+}G_{-+}(z)
&\sim1+ \frac{1}{N}\sum^\infty_{s=2} (-1)^s C_{-,0}^{(s)} C_{+,0}^{(s)} (1-z)^s {}_2F_1(s,s;2s;1-z) 
\end{align}
in this case. The extra phase factor $(-1)^s$ may require explanation;
Now we need to use a slightly different expression of operator product expansion as
\begin{align}
\mathcal{O}_+ (1) \bar{\mathcal{O}}_+ (z)  =  C_{+}^{(s)}\frac{(1 - z)^s }{|1 - z |^{2 \Delta_+}} J^{(s)} (1) + \cdots \, .
\end{align}
Then the coefficients in front of global blocks are given by
\begin{align}
C_{+}^{(s)} \langle \mathcal{O}_- (\infty) J^{(s)} (1)  \bar{\mathcal{O}}_- (0) \rangle 
=  C_{+}^{(s)} (-1)^s \langle \mathcal{O}_- (\infty)  \bar{\mathcal{O}}_- (1) J^{(s)} (0)  \rangle 
\propto (-1)^s  C_{-}^{(s)} C_{+}^{(s)}\, .
\end{align}
Here the factor $(-1)^s$ can be obtained from the coordinate dependence of three point function, which is completely fixed by conformal symmetry, see \eqref{3pt} below.
Therefore, we have constraint equations for three point functions as
\begin{align}
\frac{n-1}{n}=(\Gamma(n))^2\sum^n_{s=2}\frac{ (-1)^s  \Gamma(2s) C_{-,0}^{(s)} C_{+,0}^{(s)}}{\Gamma(s)^2\Gamma(s+n)(n-s)!}
\label{consisg-+}
\end{align}
by comparing the coefficients in front of $z^n$.

Now we have three types of constraint equation as in \eqref{consistent} and \eqref{consisg-+},
and we would like to show that the known results satisfy these equations.
At the leading order in $1/N$, the three point functions have been computed as \cite{Ammon:2011ua}
\begin{align}
	\langle \mathcal{O}_\pm (z_1) \bar{\mathcal{O}}_\pm (z_2) J^{(s)} (z_3) \rangle
	= \frac{\eta^{(s)}_\pm}{2 \pi} \frac{\Gamma (s)^2}{\Gamma(2s - 1)} 
	\frac{\Gamma(s \pm \lambda)}{\Gamma(1 \pm \lambda)} 
	\left( \frac{z_{12}}{z_{13}z_{23}} \right)^s
	\langle \mathcal{O}_\pm (z_1) \bar{\mathcal{O}}_\pm (z_2) \rangle \,  .
	\label{3pt}
\end{align}
The phase factors $\eta^{(s)}_\pm$ depends on the convention of higher spin currents, but we may set $\eta^{(s)}_+ = 1$ and  $\eta^{(s)}_- =  (-1)^s$. 
The two point function of higher spin current $J^{(s)}$ in \eqref{3pt} is (see (6.1) of \cite{Ammon:2011ua})
\begin{align}
\langle J^{(s)} (z_1) J^{(s)} (z_2) \rangle = \frac{B^{(s)}}{z_{12}^{2s}} \, , \quad
B^{(s)} = \frac{N}{2^{2s} \pi^{5/2}} \frac{\sin (\pi \lambda)}{\lambda}
\frac{\Gamma(s) \Gamma(s-\lambda) \Gamma(s+\lambda)}{\Gamma(s-\frac12)} 
\end{align}
at the leading order in $1/N$.
The coefficients $C_{\pm,0}^{(s)}$ are given by normalization independent ratios as
\begin{align}
C_{\pm,0}^{(s)} = \left.  \frac{\langle \mathcal{O}_\pm \bar{\mathcal{O}}_\pm J^{(s)} \rangle}{\langle \mathcal{O}_\pm \bar{\mathcal{O}}_\pm \rangle  \langle  J^{(s)}  J^{(s)}  \rangle ^{1/2}} \right|_{\mathcal{O}(N^{-1/2})} \, ,
\end{align}
which become
\begin{align}
C_{\pm,0}^{(s)} = \eta_{\pm}^{(s)} \sqrt{\frac{\Gamma(s)^2}{\Gamma(2s-1)}\frac{\Gamma(1 \mp \lambda)}{\Gamma (1 \pm \lambda)}
\frac{\Gamma(s \pm \lambda)}{\Gamma (s \mp \lambda)}  } \, .
\label{C3}
\end{align}
The first few coefficients are
\begin{align}
&C_{\pm , 0}^{(3)} = \eta_{\pm}^{(3)} \sqrt{\frac{1}{6} \frac{(2 \pm \lambda) (1 \pm \lambda)}{(2 \mp \lambda) ( 1 \mp \lambda) } } \, , \quad
&C_{\pm , 0}^{(4)} = \sqrt{\frac{1}{20} \frac{(3 \pm \lambda) (2 \pm \lambda) (1 \pm \lambda)}{(3 \mp \lambda)(2 \mp \lambda) ( 1 \mp \lambda) }}\label{C3f4} 
\end{align}
along with \eqref{C3f} for $s=2$.
Using these explicit expressions, we can check that the constraint equations \eqref{consistent} and \eqref{consisg-+} are indeed satisfied.%
\footnote{We have confirmed this for \eqref{consistent} with spin $s=2,3,\ldots,70$ and for \eqref{consisg-+} with all spin. }

\subsection{$1/N$ corrections}
\label{1/Ncorrections}

We would like to move to the main part of this paper.
In this subsection we derive $1/N$ corrections to three point functions by examining the equations in
\eqref{highercond}. With the help of analysis in previous sections, we have already ingredients necessary to the task.
For examples, the expansions of $G_{\pm \pm} (z)$ were given in \eqref{ExpG++} and \eqref{ExpG--} up to order $1/N^2$. Moreover, the vacuum block and the Virasoro block of $J^{(s)}$ are expanded as in \eqref{Vacexp} and \eqref{vbop}, respectively.
Using these expansions, the equations in \eqref{highercond} become
\begin{align}
\sum_{m=2}^\infty f_{\pm \pm}^{(m)}z^m = & \mathcal{V}_0^{(2)}  (z)
+ \sum_{s=3}^\infty 2 C^{(s)}_{\pm ,1} C^{(s)}_{\pm ,0} \mathcal{V}_s^{(0)} (z)+ \sum_{s=3}^\infty (C^{(s)}_{\pm ,0})^2 \mathcal{V}_s^{(1)} (z) \nonumber \\
&
 + \sum_{(s_1,s_2;s')} ( C^{(s_1,s_2;s')}_{\pm ,0} )^2 \mathcal{V}_{s ' }^{(0)} (z)
\label{refined}
\end{align}
at the order of $1/N^2$. 
Here $f_{\pm \pm}^{(m)}$ are defined in \eqref{fn++} and \eqref{fn--}.
At this order we should include the effects from higher spin currents of double trace type as
$( C^{(s_1,s_2,s')}_{\pm ,0} )^2$ in \eqref{refined} with $s' \geq 6$.

Let us examine the equations \eqref{refined} from low order terms in $z$.
There are no $z^0$ and $z^1$ order terms in the both sides.
We can see that the equality in \eqref{refined} is satisfied at the order of $z^2$ from  \eqref{vacuumN21}. 
Non-trivial constraint equations appear at the $z^3$ order as
\begin{align}
	f_{\pm \pm}^{(3)} = f^{(2)}_{\pm \pm} 
	+ 2 C_{\pm,0}^{(3)}C_{\pm,1}^{(3)} \, ,
\end{align}
where $f^{(2)}_{\pm \pm} $ comes from $\mathcal{V}_0^{(2,1)}$ in \eqref{vacuumN21}.
Solving them we find
\begin{align}
	\frac{C_{+,1}^{(3)}}{C_{+,0}^{(3)}} = - \frac{1}{2} \left( - \lambda + \frac{1}{1+ \lambda} + \frac{4}{2 + \lambda} \right) \, , \quad	\frac{C_{-,1}^{(3)}}{C_{-,0}^{(3)}} = \frac{1}{2} \left(-\lambda+\frac{1}{\lambda+1}+\frac{4}{\lambda+2}-6\right) \, .
\label{C31}
\end{align}
The $z^4$ order constraints are
\begin{align}
	f^{(4)}_{\pm \pm} = f^{(2)}_{\pm \pm} \frac{9}{10}
	+\frac{(1 \pm\lambda)^2}{8(1 \mp \lambda)^2} +\frac{(1 \pm \lambda)}{10(1 \mp \lambda)^2} + \frac{1}{50(1 \mp \lambda)^2}  + 2 C_{\pm ,0}^{(4)}C_{\pm ,1}^{(4)}
	+ 2 C_{\pm ,0}^{(3)}C_{\pm ,1}^{(3)} \frac{3}{2}\, ,  \nonumber
\end{align}
where the contribution from \eqref{vacuumN22p} starts to enter.
The constraints lead to
\begin{align}
&	\frac{C_{+,1}^{(4)}}{C_{+,0}^{(4)}}  =\frac{1}{10} \left(5 \lambda+\frac{6}{\lambda-1}-\frac{11}{\lambda+1}-\frac{20}{\lambda+2}-\frac{45}{\lambda+3}\right) \, , \nonumber \\
	&	\frac{C_{-,1}^{(4)}}{C_{-,0}^{(4)}} = \frac{1}{10} \left(-5 \lambda+\frac{6}{\lambda-1}-\frac{1}{\lambda+1}+\frac{20}{\lambda+2}+\frac{45}{\lambda+3}-60 \right) \, .\label{C41}
\end{align}
We would like to keep going to the cases with $s \geq 5$, where $f_a (s,z)$, $f_b(s,z)$, and $f_c (s,z)$ in \eqref{fb}  contribute.
For $s=5$, the conditions become
\begin{align}
	f^{(5)}_{\pm \pm} = f^{(2)}_{\pm \pm} \frac{4}{5} +  \frac{(1 \pm \lambda)^2}{4 (1 \mp \lambda)^2}
	+ \frac{(1 \pm \lambda)}{5 (1 \mp \lambda)^2} + \frac{1}{ 25 (1 \mp \lambda)^2}
	+ 2 C_{\pm ,0}^{(5)} C_{\pm ,1}^{(5)} + 2 C_{\pm ,0}^{(4)} C_{\pm ,1}^{(4)} \cdot 2 \nonumber \\
	+ 2 C_{\pm ,0}^{(3)} C_{\pm ,1}^{(3 )} \cdot \frac{12}{7} 
	+ (C^{(3)}_{\pm ,0})^2 \left[ \frac{1}{2} \frac{1 \pm \lambda}{1 \mp \lambda} + \frac{6}{7(1 \mp \lambda)}
	+ \frac{18}{49(1 - \lambda^2) } \right] \, .
\end{align}
We then find
\begin{align}
	&\frac{C^{(5)}_{+,1}}{C^{(5)}_{+,0}}
	= \frac{\lambda}{2}+\frac{25}{7 (\lambda-1)}-\frac{57}{14 (\lambda+1)}-\frac{2}{\lambda+2}-\frac{9}{2 (\lambda+3)}-\frac{8}{\lambda+4} \, , \nonumber \\
	&\frac{C^{(5)}_{-,1}}{C^{(5)}_{-,0}} =
	-\frac{\lambda}{2}+\frac{25}{7 (\lambda-1)}-\frac{43}{14 (\lambda+1)}+\frac{2}{\lambda+2}+\frac{9}{2 (\lambda+3)}+\frac{8}{\lambda+4}-10 \, \label{C51}
\end{align}
by solving the constraints.

For $s \geq 6$, the contributions from higher spin currents of double trace type should be considered. They are given by
\begin{align}
J^{(3,3;6)} \sim :J^{(3)} J^{(3)}: \, , \quad
J^{(3,4;7)} \sim :J^{(3)} J^{(4)}: 
\label{dtt67}
\end{align}
for $s=6,7$ and%
\footnote{There could be another current  $J^{(3,4;8)} \sim : J^{(3)} \partial J^{(4)}$, but it does not give any contribution as shown in appendix \ref{double}.}
\begin{align}
 J^{(4,4;8)} \sim :J^{(4)} J^{(4)}: \, , \quad
 J^{(3,5;8)} \sim :J^{(3)} J^{(5)}: \, , \quad
 J^{(3,3;8)} \sim :J^{(3)} \partial^2 J^{(3)}: 
 \label{dtt8}
\end{align}
for $s=8$. Their precise forms are fixed such as to be primary in the sense of Virasoro algebra as derived in appendix \ref{double}.%
\footnote{From the decomposition by Virasoro conformal blocks as in \eqref{AbsCPWE}, we can read off three point functions among primary operators including intermediate one $\mathcal{A}_p$ by construction. For $s \leq 5$, only $J^{(s)}$ starts to contribute as the intermediate operator at the $z^s$ order, so we do not need to worry about if the operator is primary or not. However, for $s \geq 6$, there are degeneracies among $J^{(s)}$ and $J^{(s_1 ,s_2 ; s)}$, and the $1/N$ corrections $C^{(s)}_{\pm , 1}$ could be read off once we have the information of $C_{\pm ,0}^{(s_1,s_2 ; s)}$, see \eqref{refined}. Since we compute $C_{\pm ,0}^{(s_1,s_2 ; s)}$ by hand as explained in appendix \ref{double}, we have to explicitly construct primary operators of double trace type. We  only  need the leading order expressions, so it is enough to use commutation relations surviving in the large $c$ limit as in \eqref{commutator} and higher spin charges at the leading order as in \eqref{hsc}.}
Once we have the expressions of these currents, we can obtain the coefficients  
$(C^{(s_1,s_2;s')}_{\pm ,0})^2$, which are defined as
\begin{align}
(C^{(s_1,s_2;s')}_{\pm, 0 })^2 = \left. \frac{\langle \mathcal{O}_\pm \bar{\mathcal{O}}_\pm J^{(s_1,s_2;s')} \rangle^2}{\langle \mathcal{O}_\pm \bar{\mathcal{O}}_\pm \rangle^2\langle  J^{(s_1,s_2;s')} J^{(s_1,s_2;s')} \rangle} \right|_{\mathcal{O}(N^{-2})} \, .
\label{dtt3pt}
\end{align}
In appendix \ref{double} we also compute the three point functions $\langle \mathcal{O}_\pm \bar{\mathcal{O}}_\pm J^{(s_1,s_2;s')} \rangle $ and the two point functions $\langle  J^{(s_1,s_2;s')} J^{(s_1,s_2;s')} \rangle$ for the currents in \eqref{dtt67} and \eqref{dtt8} at the leading order in $1/N$.

Utilizing these results, we obtain $1/N$ corrections to three point functions with single trace currents of $s=6,7,8$.
The constraint equations for $s=6$ are
\begin{align}
	f^{(6)}_{\pm \pm} &= f^{(2)}_{\pm \pm} \frac{5}{7} +  \frac{7(1 \pm \lambda)^2}{20 (1 \mp \lambda)^2}
	+ \frac{39(1 \pm \lambda)}{140 (1 \mp \lambda)^2} + \frac{109}{ 1960 (1 \mp \lambda)^2}
	+ 2 C_{\pm ,0}^{(6)} C_{\pm ,1}^{(6)}+ 2 C_{\pm ,0}^{(5)} C_{\pm ,1}^{(5)} \cdot \frac{5}{2} \nonumber \\ &+ 2 C_{\pm ,0}^{(4)} C_{\pm ,1}^{(4)} \cdot \frac{25}{9} 
	+ 2 C_{\pm ,0}^{(3)} C_{\pm ,1}^{(3 )} \cdot \frac{25}{14} 
	+ (C^{(3)}_{\pm ,0})^2 \left[ \frac{5}{4} \frac{1 \pm \lambda}{1 \mp \lambda} + \frac{15}{7(1 \mp \lambda)}
	+ \frac{90}{98(1 - \lambda^2) }  \right] \nonumber \\
	&+ (C^{(4)}_{\pm ,0})^2 \left[ \frac{1}{2} \frac{1 \pm \lambda}{1 \mp \lambda} + \frac{4}{3(1 \mp \lambda)}
	+ \frac{8}{9(1 - \lambda^2) } \right]  + (C^{(3,3;6)}_{\pm ,0})^2 \, ,
	\label{conditions6}
\end{align}
where the effect of $J^{(3,3;6)}$ in \eqref{dtt67} enters.
Solving these equations we find
\begin{align}
	\frac{C^{(6)}_{+,1}}{C^{(6)}_{+,0}} 
	&= \frac{\lambda}{2}-\frac{5}{3 (\lambda-2)}+\frac{1315}{84 (\lambda-1)}-\frac{1357}{84 (\lambda+1)}-\frac{1}{3 (\lambda+2)}-\frac{9}{2 (\lambda+3)} -\frac{8}{\lambda+4}  \nonumber \\€€& -\frac{25}{2 (\lambda+5)} \, , 
	\label{C61} \\
	\frac{C^{(6)}_{-,1}}{C^{(6)}_{-,0}} &=
	-\frac{\lambda}{2}-\frac{5}{3 (\lambda-2)}+\frac{1315}{84 (\lambda-1)}-\frac{1273}{84 (\lambda+1)}+\frac{11}{3 (\lambda+2)}+\frac{9}{2 (\lambda+3)}  +\frac{8}{\lambda+4}\nonumber \\€€&+\frac{25}{2 (\lambda+5)}-15 \, .\nonumber 
\end{align}
For spin 7,  another double trace operator $J^{(3,4;7)}$ in \eqref{dtt67} should be considered as
\begin{align}
	f^{(7)}_{\pm \pm} &= f^{(2)}_{\pm \pm} \frac{9}{14} +  \frac{17(1 \pm \lambda)^2}{40 (1 \mp \lambda)^2}
	+ \frac{846 (1 \pm \lambda)}{2520 (1 \mp \lambda)^2} + \frac{131}{ 1960 (1 \mp \lambda)^2}+ 2 C_{\pm ,0}^{(7)} C_{\pm ,1}^{(7)} 
	+ 2 C_{\pm ,0}^{(6)} C_{\pm ,1}^{(6)} \cdot 3 \nonumber \\ &+ 2 C_{\pm ,0}^{(5)} C_{\pm ,1}^{(5)} \cdot \frac{45}{11} + 2 C_{\pm ,0}^{(4)} C_{\pm ,1}^{(4)} \cdot \frac{10}{3} 
	+ 2 C_{\pm ,0}^{(3)} C_{\pm ,1}^{(3 )} \cdot \frac{25}{14} 
	+ (C^{(3)}_{\pm ,0})^2  \frac{(3467 \pm 42 \lambda (89 \pm 24 \lambda) )}{490(1 - \lambda^2)}  \nonumber \\
	&+ (C^{(4)}_{\pm ,0})^2  \frac{(7 \pm 3 \lambda)}{6(1 - \lambda^2)} 
	+ (C^{(5)}_{\pm ,0})^2  \frac{(31 \pm 11 \lambda)}{242 (1 - \lambda^2)} 
	+ (C^{(3,3;6)}_{\pm ,0})^2 \cdot 3 +  (C^{(3,4;7)}_{\pm ,0})^2 \, ,
	\label{conditions7}
\end{align}
which lead to
\begin{align}
	&\frac{C^{(7)}_{+,1}}{C^{(7)}_{+,0}}
	= \frac{\lambda}{2}-\frac{490}{33 (\lambda-2)}+\frac{8183}{132 (\lambda-1)}-\frac{8249}{132 (\lambda+1)}+\frac{424}{33 (\lambda+2)}-\frac{9}{2 (\lambda+3)} \nonumber \\ & \qquad -\frac{8}{\lambda+4}  -\frac{25}{2 (\lambda+5)}-\frac{18}{\lambda+6}\, , \label{C71}\\
	&\frac{C^{(7)}_{-,1}}{C^{(7)}_{-,0}} =
	-\frac{\lambda}{2}-\frac{490}{33 (\lambda-2)}+\frac{8183}{132 (\lambda-1)}-\frac{8117}{132 (\lambda+1)}+\frac{556}{33 (\lambda+2)}+\frac{9}{2 (\lambda+3)}\nonumber \\ & \qquad +\frac{8}{\lambda+4} +\frac{25}{2 (\lambda+5)}+\frac{18}{\lambda+6}-21 \, . \nonumber 
\end{align}
The constraint equations for $C_{\pm,1}^{(8)}$ are
\begin{align}
&f^{(8)}_{\pm \pm} = f^{(2)}_{\pm \pm} \frac{7}{12} + \frac{(1 \pm \lambda)^2}{ (1 \mp \lambda)^2} \frac{2687}{5600} + \frac{ (1 \pm \lambda)}{(1 \mp \lambda)^2} \frac{263}{700} + \frac{1}{(1 \mp \lambda)^2} \frac{1879}{25200} 
+ 2 C_{\pm ,0}^{(8)} C_{\pm ,1}^{(8)} \nonumber  \\&+ 2 C_{\pm ,0}^{(7)} C_{\pm ,1}^{(7)} \frac{7}{2}  + 2 C_{\pm ,0}^{(6)} C_{\pm ,1}^{(6)} \frac{147}{26}  + 2 C_{\pm ,0}^{(5)} C_{\pm ,1}^{(5)} \frac{245}{44} + + 2 C_{\pm ,0}^{(4)} C_{\pm ,1}^{(4)} \frac{245}{66} + 2 C_{\pm ,0}^{(3)} C_{\pm ,1}^{(3)} \frac{7}{4} \nonumber \\& + (C_{\pm,0}^{(3)})^2 \left( \frac{387 \pm 418 \lambda + 113 \lambda^2}{40 (1 - \lambda^2)}\right) + (C_{\pm,0}^{(4)})^2 \left( \frac{7 (47977 \pm 41162 \lambda + 8833  \lambda^2)}{21780 (1 - \lambda^2)}\right)\\& + (C_{\pm,0}^{(5)})^2 \left( \frac{7 (31 \pm 11 \lambda)^2}{484 (1 - \lambda^2)}\right) + (C_{\pm,0}^{(6)})^2 \left( \frac{ (43 \pm 13 \lambda)^2}{338 (1 - \lambda^2)}\right) + (C_{\pm ,0}^{(3,3;6)}) ^2 \frac{147}{26} + (C_{\pm ,0}^{(3,4;7)}) ^2 \frac{7}{2} \nonumber \\& +   (C_{\pm ,0}^{(4,4;8)}) ^2 +   (C_{\pm ,0}^{(3,5;8)}) ^2+   (C_{\pm ,0}^{(3,3;8)}) ^2 \, .\nonumber 
\end{align}
Here we have taken care of double trace operators $J^{(4,4;8)}$, $J^{(3,5;8)}$, and $J^{(3,3;8)}$ in 
\eqref{dtt8}.
We then have 
\begin{align}
&\frac{C^{(8)}_{+,1}}{C^{(8)}_{+,0}}
= \frac{\lambda}{2}+\frac{525}{143 (\lambda-3)}-\frac{12572}{143 (\lambda-2)}+\frac{101311}{429 (\lambda-1)} -\frac{203051}{858 (\lambda+1)}+\frac{12286}{143 (\lambda+2)} \nonumber \\ & \qquad  -\frac{2337}{286 (\lambda+3)} -\frac{8}{\lambda+4}-\frac{25}{2 (\lambda+5)}-\frac{18}{\lambda+6}-\frac{49}{2 (\lambda+7)}\, , \label{C81}\\
&\frac{C^{(8)}_{-,1}}{C^{(8)}_{-,0}} =
-\frac{\lambda}{2}+\frac{525}{143 (\lambda-3)}-\frac{12572}{143 (\lambda-2)}+\frac{101311}{429 (\lambda-1)}-\frac{202193}{858 (\lambda+1)}+\frac{12858}{143 (\lambda+2)} \nonumber \\ & \qquad+\frac{237}{286 (\lambda+3)}+\frac{8}{\lambda+4}+\frac{25}{2 (\lambda+5)}+\frac{18}{\lambda+6}+\frac{49}{2 (\lambda+7)}-28  \nonumber 
\end{align}
as solutions to the constraint equations.

Since the three point functions were already obtained with finite $N,k$ in \cite{Bais:1987zk,Ahn:2011by,Ahn:2013sua} for $s=3,4,5$, they can be compared to our results in principle. Instead of doing so, we utilize a simpler relation, which is on  the ratio of three point functions  (see (4.52) of \cite{Ahn:2013sua})
\begin{align}
\frac{\langle  \mathcal{O}_+ \bar{\mathcal{O}}_+ J^{(s)} \rangle}{\langle \mathcal{O}_- \bar{\mathcal{O}}_-   J^{(s)} \rangle}
= (-1)^s \frac{(k+N+1)}{(k+N)} \prod_{n=1}^{s-1} \left[ \frac{n k + (n+1) N + n}{n k + (n-1) N}\right] \, . \label{Ahnconjecture}
\end{align}
The relation was derived for $s=2,3,4,5$ by using the explicit results  and conjectured for generic $s$ based on them.
The expression up to the $1/N$ order becomes
\begin{align}
\frac{\langle\mathcal{O}_+  \bar{\mathcal{O}}_+  J^{(s)} \rangle}{\langle \mathcal{O}_- \bar{\mathcal{O}}_-  J^{(s)} \rangle}
&= \frac{C_{+,0}^{(s)} + \frac{1}{N} C_{+,1}^{(s)}}{C_{-,0}^{(s)} + \frac{1}{N} C_{-,1}^{(s)}} + \mathcal{O} (N^{-2} ) \nonumber  \\
&= (-1)^s \prod_{n=1}^{s-1} \left( \frac{n+ \lambda}{n - \lambda}\right) 
\left[ 1 + \frac{1}{N} \left( \lambda + \sum_{m=1}^{s-1} \frac{m \lambda}{m+ \lambda}\right) + \mathcal{O}(N^{-2})\right] \, .
\label{consistency}
\end{align}
Thus, at the leading order in $1/N$, we have
\begin{align}
\frac{C^{(s)}_{+,0}}{C^{(s)}_{-,0}} = (-1)^s \prod_{n=1}^{s-1} \left( \frac{n+\lambda}{n-\lambda} \right) \, .
\end{align}
We can easily check that \eqref{C3} satisfy this condition.
The relation in \eqref{consistency} at the next leading order in $1/N$ implies
\begin{align}
\frac{C^{(s)}_{+,1}}{C^{(s)}_{+,0}} - \frac{C^{(s)}_{-,1}}{C^{(s)}_{-,0}} 
= \lambda + \sum_{m=1}^{s-1} \frac{m \lambda}{m + \lambda} \, . \label{check}
\end{align}
We have confirmed our results (and the conjectured relation in \eqref{consistency}) by showing
that our results on $C^{(s)}_{\pm,1}$ for $s=3,\ldots,8$ satisfy this equation.

Before ending this section, we would like to make comments on normalized three point functions
\begin{align}
C_{\pm}^{(2)} = 
\frac{\langle \mathcal{O}_\pm \bar{ \mathcal{O}}_\pm J^{(2)} \rangle}{\langle \mathcal{O}_\pm \bar{ \mathcal{O}}_\pm  \rangle \langle    J^{(2)}  J^{(2)} \rangle^{1/2}}  
\end{align}
with the energy momentum tensor $T \propto J^{(2)}$.
They do not appear in the decomposition of Virasoro conformal blocks but can be fixed by the conformal Ward identity as
\begin{align}
C_{\pm}^{(2)} = \frac{1}{N^{1/2}} \left(C_{\pm ,0}^{(2)} + \frac{1}{N} C_{\pm,1}^{(2)} + \mathcal{O} (N^{-2}) \right) =  \sqrt{ \frac{2 h_\pm^2 }{c}} \, .
\end{align}
In particular, they lead to \eqref{C3f} and
\begin{align}
2 C_{\pm ,0}^{(2)} C_{\pm,1}^{(2)}  = f^{(2)}_{\pm \pm}
\end{align}
with \eqref{cwexp}, or equivalently
\begin{align}
\frac{C_{+ ,1}^{(2)}}{C_{+ ,0}} = \frac{\lambda}{2} - \frac{1}{2 (\lambda+1)} \, , 
\quad 
\frac{C_{-,1}^{(2)}}{C_{-,0}} =-\frac{\lambda}{2}+\frac{1}{2 (\lambda+1)}-1 \, .
\label{C21}
\end{align}
As a consistence check, we can show that they satisfy \eqref{check} as well.

\section{Conclusion and open problems}
\label{conclusion}

We have developed a new method to compute three point functions of two scalar operators and a higher spin current \eqref{3pt0} in 2d W$_N$ minimal model.
This model can be described by the coset \eqref{coset} with two parameters $N,k$, and we analyze it in $1/N$ expansion in terms of 't Hooft parameter $\lambda = N/(N+k)$ in \eqref{thooft}.
We decompose scalar four point functions $G_{\pm\pm}(z)$ in \eqref{G++0}, \eqref{G--0} and $G_{-+}(z)$ in \eqref{G-+0} by Virasoro  conformal blocks.
The four point functions were computed exactly with finite $N,k$  in \cite{Papadodimas:2011pf}, and Virasoro  conformal blocks can be obtained including $1/N$ corrections, say, by analyzing Zamolodchikov's recursion relation \cite{Zamolodchikov:1985ie}. Solving the constraint equations from the decomposition, we can obtain three point functions including $1/N$ corrections.
At the leading order in $1/N$, we can easily reproduce the known results in \cite{Ammon:2011ua} because Virasoro conformal blocks reduce to global blocks in this case.
At the next leading order, we have obtained $1/N$ corrections to the three point functions up to spin $8$. Previously exact results were known for $s=3,4,5$ in \cite{Bais:1987zk,Ahn:2011by,Ahn:2013sua}, and our findings for $s=6,7,8$ are new. We have confirmed our results by checking that the conjectured relation in \eqref{check} is satisfied.

We have evaluated $1/N$ corrections only up to spin $8$ case because of the following two obstacles.  One comes from $1/c$ corrections to Virasoro conformal blocks. Up to the required order in $1/c$, closed forms can be obtained, for instance, by following the method in \cite{Fitzpatrick:2016mtp} except for $f_c (s,z)$ in \eqref{vbop2}.
In \eqref{fb} (or in \cite{Fitzpatrick:2016mtp}), the function $f_c(s,z)$ is given up to the order $z^{5+s}$, but we need the term at order $z^{6+s}$ with $s=3$ for spin 9 computation.
We have not tried to do so, but it should be possible to obtain the terms at higher orders in $z$ without a lot of efforts.
Another is related to the contributions from higher spin currents of double trace type as analyzed in appendix \ref{double}. In order to obtain primary operators of this type, we have used commutation relations in \eqref{commutator}, which are borrowed from \cite{Gaberdiel:2012yb}. For spin 9, a current of the form $J^{(3,6;9)} \sim :J^{(3)} J^{(6)}:$ would give some contributions. However, in order to find its primary form, we need the commutation relation between $W,Y$, which is currently not available. At the order in $1/c$ which do not vanish at $c \to \infty$, we can derive the commutation relations involving more higher spin currents, for instance, from dual Chern-Simons description as in \cite{Henneaux:2010xg,Campoleoni:2010zq,Gaberdiel:2011wb,Campoleoni:2011hg}.
The computation is straightforward but might be tedious.
In any case, it is definitely possible to obtain the $1/N$ corrections of three point functions for $s \geq 9$, and it is desired to have expressions for generic $s$.

There are many open problems we would like to think about.
Because of the simplicity of our method, it is expected to be applicable to more generic cases. 
For example,  it is worth generalizing the current analysis to  supersymmetric cases.
Recently, it becomes possible to discuss relations between 3d higher spin theory and superstrings by introducing extended supersymmetry to the duality by \cite{Gaberdiel:2010pz}. 
Higher spin holography with $\mathcal{N}=3$ supersymmetry has been developed in a series of works  \cite{Creutzig:2013tja,Creutzig:2014ula,Hikida:2015nfa}, while large or small $\mathcal{N}=4$ supersymmetry has been utilized through the well-studied holography with symmetric orbifold in \cite{Gaberdiel:2013vva,Gaberdiel:2014cha}.
Previous works on the subject may be found in \cite{Ahn:2015rma,Ahn:2015gxa,Ahn:2017dqo}.
As mentioned in introduction, the main motivation to examine $1/N$ corrections in 2d W$_N$ minimal model is to learn quantum effects in dual higher spin theory. 
We would like to report on our recent progress  in a separate publication \cite{Hikida:2017ehf}.

\subsection*{Acknowledgements}

We are grateful to Changhyun Ahn, Pawel Caputa, Takahiro Nishinaka, Volker Schomerus and Yuji Sugawara for useful discussion. 
YH would like to thank the organizers of ``Universit\"{a}t
Hamburg - Kyoto University Symposium" and the workshop ``New ideas on higher spin gravity and holography'' at Kyung Hee University, Seoul for their hospitality.
The work of YH is supported by JSPS KAKENHI Grant Number 16H02182.

\appendix

\section{Recursion relations and Virasoro conformal blocks}
\label{Zamolodchikov}

In this appendix we derive the expressions of Virasoro conformal blocks in expansions of $1/c$ and $z$ by solving Zamolodchikov recursion relation in  \cite{Zamolodchikov:1985ie}, and we compare our results to those previous obtained especially in \cite{Fitzpatrick:2016mtp}. 
We decompose a four point function  by Virasoro conformal blocks $\mathcal{F}(c,h_i,h_p,z)$ as in \eqref{AbsCPWE}.
In the following we set $h_1 = h_2$ and $h_3 = h_4$.
The recursion relation for Virasoro conformal blocks is \cite{Zamolodchikov:1985ie}
\begin{align}
	\mathcal{F} (c,h_i,h_p,z) = & z^{h_p} {}_2 F_1 (h_p  , h_p  ; 2 h_p ;z) \nonumber \\
	& +\sum_{m \geq 1 ,n \geq 2}^\infty \frac{R_{mn} (h_i,h_p)}{c - c_{mn} (h_p)} \mathcal{F}(c_{mn}(h_p),h_i,h_p + mn ,z) \, .
\end{align}
Here the poles for $c$ are located at $c = c_{mn}(h_p)$ with
\begin{align}
	c_{mn} (h_p) = 13 - 6 \left( t_{mn} (h_p) ^{-1} + t_{mn} (h_p) \right) \, ,
\end{align}
where
\begin{align}
	t_{mn} (h_p) = \left( 2 h_p + mn - 1 + \sqrt{4h_p (h_p + mn - 1) + (m-n)^2}\right)/(n^2 -1) \, .
\end{align}
The residua are 
\begin{align}
	R_{mn} (h_i ,h_p) = A_{mn} (h_p) P_{mn} (h_i , h_p) \, ,
\end{align}
where
\begin{align}
	&P_{mn} (h_i ,h_p) = \prod_{j,k} \left(2 l_1 - \frac{l_{jk}}{2} \right)
	\left(2 l_3 - \frac{l_{jk}}{2} \right)
	\left( \frac{l_{jk}}{2} \right) ^2 \, , \nonumber \\
	&A_{mn}(h_p) =  \frac{- 12(t^{-1}_{mn} - t_{mn})}{(m^2-1)t_{mn}^{-1}-(n^2-1)t_{mn}}
	\prod_{a,b} \frac{1}{l_{ab}} \, , \\
	&l_{jk } (m,n,h_p) = (j - k t_{mn}) t_{mn}^{-1/2} \, , \quad
	l_i (m,n,h_i,h_p) = (h_i + l^2_{11}/4)^{1/2} \, . \nonumber
\end{align}
The sum is taken over $j=-m+1,-m+3,\ldots,m-1, k=-n+1,-n+3,\ldots,n-1,a=-m+1,-m+2,\ldots,m,b=-n+1,-n+2,\ldots,n$ without $(a,b)=(0,0), (m,n)$.

For our purpose, it is enough to obtain first several terms of Virasoro blocks in $z$ expansion, and we obtain them  by following the strategy of \cite{Perlmutter:2015iya}, see also \cite{Beccaria:2015shq}. We decompose Virasoro conformal blocks by global blocks as
\begin{align}
	\mathcal{F} (c,h_i,h_p,z) = z^{h_p} \sum_{q=0}^\infty \chi_q (c,h_i ,h_p) z^q
	{}_2 F_1 (h_p + q , h_p + q ; 2 (h_p + q) ; z) \, .
	\label{globaldec}
\end{align}
The generic expressions of $\chi_q$ are given in (2.28) of \cite{Perlmutter:2015iya}.
With $h_1=h_2$ and $h_3=h_4$, it can be shown that $\chi_q = 0$ for odd $q$.
The explicit expressions for $q=2, 4 ,6$ can be found in (C.1) of the paper as
\begin{align}
	&\chi_2 (c ,h_p) = \gamma_{12} (c,h_p) \, , \nonumber \\
	&\chi_4 (c , h_p) = \gamma_{14} (c,h_p) + \gamma_{22} (c,h_p)  + \gamma_{12}(c,h_p) \gamma_{12} (c_{12} (h_p,h_p+2) ) \, , \nonumber\\
	&\chi_6 (c,h_p) = \gamma_{16} (c,h_p) + \gamma_{23} (c,h_p) + \gamma_{32} (c,h_p) 
	+ \gamma_{12} (c,h_p) \gamma_{14} (c_{12} (h_p) , h_p + 2)  \label{chiqfew} \\
	& \quad + \gamma_{12} (c , h_p) 
	\gamma_{22} (c_{12}(h_p),h_p + 2) + \gamma_{14} (c,h_p) \gamma_{12} (c_{14}(h_p),h_p+4) 
	\nonumber\\ & \quad  + \gamma_{22} (c , h_p)   \gamma_{12} (c_{22}(h_p),h_p + 4) + \gamma_{12} (c , h_p) \gamma_{12} (c_{12}(h_p),h_p + 2)   
	\gamma_{12} (c_{12}(h_p+2),h_p + 4)   \nonumber
\end{align}
with
\begin{align}
	\gamma_{mn} (c,h_p) = \frac{R_{mn} (h_i,h_p)}{c - c_{mn} (h_p)} \, .
\end{align}
Inserting these expressions into \eqref{globaldec}, we can obtain the Virasoro conformal blocks up to the order of $z^{h_p +7}$.

Let us start from vacuum block. As discussed in the main context, we need its expression up to the $1/c^2$ order.  
For $h_p = 0$ the coefficients $\chi_q$ can be found in (2.15) of \cite{Perlmutter:2015iya},
and they are expended in $1/c$ as 
\begin{align}
	&\chi_2 (c , h_i ,0) = \frac{2h_1 h_3}{c} \, , \nonumber\\
	&\chi_4 (c , h_i ,0) = \frac{2 (5 h_1^2 +h_1)(5 h_2^2 + h_2 )}{25 c^2} + \mathcal{O} (c^{-3}) \, ,\\
	&\chi_6 (c , h_i ,0) = \frac{(14 h_1^2 + h_1 ) (14 h^2_3 + h_3)}{4410 c^2 }+ \mathcal{O} (c^{-3}) \, . \nonumber
\end{align}
Note that there is no $1/c$-correction to $\chi_2 (c , h_i ,0)$. 
Using
\begin{align}
&	{}_2 F_1 (4,4;8;z) = 1 + 2 z + \frac{25}{9}z^2 + \frac{10}{3} z^3 + \mathcal{O}(z^4) \, , \nonumber \\
&	{}_2 F_1 (6,6;12;z) = 1 + 3 z + \mathcal{O}(z^2) \, , 
\end{align}
and \eqref{globaldec},
we find \eqref{vbexp}
with $k_a(z)$, $k_b(z)$, and $k_c(z)$  in \eqref{kb} but up to the order $z^7$.

We would like to compare the expressions to eq.~(3.14) in \cite{Fitzpatrick:2016mtp}.
Firstly, there is no contribution like $k_a(z)$.%
\footnote{It seems that the authors of \cite{Fitzpatrick:2016mtp} did not consider this type of contribution because it is not new but essentially given by the square of $1/c$ order term in \eqref{vbexp}.}
Secondly, our $k_b(z)$ is twice of the corresponding one in  \cite{Fitzpatrick:2016mtp}.
Finally, we can see that $k_c(z)$ reproduces their expressions.
In conclusion, we find very similar but different results.
After carefully repeated the analysis, say, in \cite{Fitzpatrick:2016mtp}, we obtain
\begin{align}
	& k_a (z)=   2 z^4 ({}_2 F_1 (2,2;4;z))^2 \, ,  \nonumber  \\
	& k_b (z)=  \frac{72   }{ z^{2}} 
	( (z-2 )z \log (1-z) + 2 (1-z) \log ^2 (1-z) - 4 z^2 )  \, ,	\label{vacuumN22} \\
	& k_c (z)=  \frac{12}{ z^{2} } 
(12 (z-2) z \text{Li}_2 (z) + 16 z^2 + 6 (z-1)^2 \log ^2 (1-z) + (z-2) z \log (1-z) )  \, . \nonumber
\end{align}
This version matches the above expressions in $z$ expansion.
Using these closed form results, we can go to more higher orders in $z$ as in \eqref{kb}.

Let us move to the case with non-trivial $h_p$, where expressions are needed up to the $1/c$ order.
Using the expressions of $\chi_q$ in \eqref{chiqfew}, we obtain
\begin{align}
	&	\chi_2 (c,h_p) = \frac{1}{c} \left( \frac{h_p^2 (h_p-1)^2}{2 ( 2h_p + 1)^2} + (h_1 + h_3 ) 
	\frac{h_p (h_p-1)}{2 h_p + 1} + 2 h_1 h_3 
	\right) + \mathcal{O}(c^{-2}) \, , \nonumber \\
	&	\chi_4 (c,h_p) = \frac{1}{c} \left( \frac{(h_p-1)^2 h_p^3 (h_p+3)}{80  (2 h_p+1) (2 h_p+3)^2 (2 h_p+5)}\right. \\
	& \quad \quad\left.+ (h_1 + h_3 ) 
	\frac{(h_p-1) h_p^2 (h_p+3) }{20  (2 h_p+1) (2 h_p+3) (2 h_p+5)} + h_1 h_3 \frac{ h_p (h_p+3) }{5  (2 h_p+1) (2 h_p+5)}
	\right) + \mathcal{O}(c^{-2}) \, .\nonumber
\end{align}
With the expansions of hypergeometric function in $z$ such as
\begin{align}
	&{}_2 F_1 (h_p + 2, h_p + 2;2 h_p + 4;z) = 1 +\frac{2 + h_p}{2} z + \frac{(2 + h_p)(3 + h_p)^2}{4 (5 + 2h_p)} z^2 \nonumber \\
	& \qquad \qquad \qquad  \qquad \qquad \qquad \qquad  +\frac{(h_p+2) (h_p+3) (h_p+4)^2 }{24 (2 h_p +5)} z^3 + \mathcal{O}(z^4) \, , \\
	&{}_2 F_1 (h_p + 4, h_p + 4;2 h_p + 8;z) = 1 + \frac{4 + h_p}{2} z + \mathcal{O}(z^2)  \, , \nonumber
\end{align}
we find \eqref{vbop0}, where the functions $f_a (h_p , z)$, $f_b (h_p , z)$, and $f_c (h_p , z)$ are given by \eqref{fb} but with $s=h_p$. 
These were analyzed in \cite{Fitzpatrick:2016mtp} and, in particular, closed forms were obtained for $f_a (h_p ,z)$ and $f_b (h_p , z)$ as
\begin{align}
&f_a (h_p ,z) = - 12 z^{h_p -1} {}_2 F_1 (h_p,h_p; 2h_p ; z) (2 z + (2 z + (2 - z) \log (1 - z) )) \, , \nonumber \\
&f_b (h_p , z) = 12 h_p z^{h_p}  \left(\, _2F_1(h_p,h_p; 2 h_p;z) \left(\log (1-z) \left(z^{-1}-1\right)+1\right)
 \right.  \\ 
 &\qquad \qquad \qquad \qquad \qquad \qquad \qquad  \left. + \tfrac12 \log (1-z)  \, _2F_1(h_p,h_p; 2 h_p+1;z) \right)  \nonumber  \, .
\end{align}
Our results match with their findings in this case.

\section{Higher spin currents of double trace type}
\label{double}

In this appendix, we analyze higher spin currents of double trace type with $s' = 6,7,8$ in \eqref{dtt67} and \eqref{dtt8}.
We first present basics on higher spin algebra, which are needed to obtain the precise expressions of these currents primary with respect to Virasoro algebra.
We  then derive the three and two point functions of these currents, which are used to  obtain $(C^{(s_1,s_2;s')}_{\pm,0})^2$ in \eqref{dtt3pt}.

\subsection{Higher spin algebra}

In order to find out higher spin currents of double trace type, which are primary to Virasoro algebra,
we utilize  commutation relations among higher spin currents given in \cite{Gaberdiel:2012yb} (see also \cite{Gaberdiel:2011wb,Campoleoni:2011hg,Gaberdiel:2012ku}).
The currents are denoted as $W,U,X,Y$, which are proportional to $J^{(s)}$ with $s=3,4,5,6$. 
In order to obtain the leading order expression $(C^{(s_1,s_2;s')}_{\pm,0})^2$, we only need commutation relations up to the terms vanishing at $c \to \infty$ as%
\footnote{Here we have changed some signs, see, e.g., footnote 6 of \cite{Creutzig:2015hta}.
The changes here are associated with redefinitions as $W \to i W$, $U \to - U$, $X \to -i X$, and $Y$ untouched.}
\begin{align}
	& [L_m , L_n] = (m - n) L_{m+n} + \frac{c}{12} m (m^2 -1) \delta_{m+n} \, , \quad [L_m , W_n] = (2m - n) W_{m+n} \, , \nonumber \\
	&[L_m , U_n] = (3m - n) U_{m+n} \, , \quad [L_m , X_n] = (4 m - n) X_{m+n} \, , \quad
	[L_m , Y_n] = (5m - n) Y_{m+n} \, , \nonumber \\
	& [W_m , W_n] = 2 (m - n) U_{m+n} - \frac{N_3}{12} (m - n) (2 m^2 + 2 n^2 - m n - 8) L_{m+n} \nonumber \\
	&   - \frac{N_3 c}{144} m (m^2 - 1) (m^2 - 4) \delta_{m+n} \, , \nonumber \\
	&[W_m , U_n] = (3m - 2n) X_{m+n} - \frac{N_4}{15 N _ 3} (n^3 - 5 m^3 - 3 m n^2 + 5 m^2 n - 9 n + 17 m) W_{m+n} \, , \nonumber \\
	&[U_m , U_n] = 3 (m-n) Y_{m+n} - n_{44} (m-n) (m^2 - mn + n^2 - 7) U_{m+n} \label{commutator} \\
	&   - \frac{N_4}{360} (m-n)(108 - 39 m^2 + 3 m^4 + 20 m n - 2 m^3 n - 39 n^2 + 4m^2 n^2 - 2 m n^2 + 3 n^4   ) L_{m+n} \nonumber \\
	& - \frac{c N_4 }{4320} m (m^2 -1) (m^2 -4) (m^2 - 9) \delta_{m+n} \, , \nonumber \\
	&[W_m , X_n] = (4 m - 2n) Y_{m+n} + \frac{1}{56} \frac{N_5}{N_4} (28 m^3 - 21 m^2 n + 9 m n^2 - 2 n^3 - 88 m + 32 n) U_{m+n} \, , \nonumber \\
	&[X_m , X_n] = - \frac{c N_5}{241920} m (m^2 -1) (m^2 - 4) (m^2 - 9) (m^2 - 16) \delta_{m+n} + \cdots \, .\nonumber
\end{align}
The constants are
\begin{align}
	&N_3 = \frac{1}{5} (\lambda ^2 - 4) \, , \quad N_4 = - \frac{3}{70} (\lambda^2 - 4) (\lambda^2 - 9) \, , \\
	&N_5 =  \frac{1}{105} (\lambda^2 -4) (\lambda^2 - 9) (\lambda^2 - 16) \, , \quad
	n_{44} = \frac{1}{30} (\lambda^2 - 19)  \nonumber
\end{align}
in the current notation.

With the conventions,  higher spin charges are given by 
\begin{align}
	&L_0 | \mathcal{O}_\pm \rangle = h | \mathcal{O}_\pm \rangle  \, , \quad
	W_0 | \mathcal{O}_\pm \rangle = w | \mathcal{O}_\pm \rangle  \, , \quad
	U_0 | \mathcal{O}_\pm \rangle = u | \mathcal{O}_\pm \rangle  \, , \\
	&X_0 | \mathcal{O}_\pm \rangle = x | \mathcal{O}_\pm \rangle  \, , \quad
	Y_0 | \mathcal{O}_\pm \rangle = y | \mathcal{O}_\pm \rangle  \, .  \nonumber 
\end{align}
Here $| \mathcal{O}_\pm \rangle  \equiv \mathcal{O}_\pm (0) | 0 \rangle$ and 
\begin{align}
\label{hsc}
	&h = \frac12 (1 \pm \lambda) \, , \quad
	w = \pm \frac16 (2 \pm \lambda) (1 \pm \lambda) \, , \quad
	u = \frac{1}{20} (3 \pm \lambda) (2 \pm \lambda) (1 \pm \lambda) \, ,\\
	&x= \pm \frac{1}{70} (4 \pm \lambda) (3 \pm \lambda) (2 \pm \lambda) (1 \pm \lambda) \, ,\quad
	y = \frac{1}{252} (5 \pm \lambda)  (4 \pm \lambda) (3 \pm \lambda) (2 \pm \lambda) (1 \pm \lambda) \nonumber
\end{align}
at the leading order in $1/c$.

\subsection{Three and two point functions}
\label{23pt}

We start from spin 6 current $J^{(3,3;6)}  \sim :J^{(3)} J^{(3)}: $ in \eqref{dtt67}.
Let us assume the form as
\begin{align}
	J^{(3,3;6)} (0)| 0 \rangle = J^{(3,3;6)}_{-6} | 0 \rangle
	= (W_{-3} W_{-3} + a U_{-6} + b L_{-6}) | 0 \rangle  \, .
\end{align}
Then the coefficients $a,b$ are fixed by the condition $L_{1} J^{(3,3;6)}_{-6} | 0 \rangle = 0$
as
\begin{align}
	a = - \frac{10}{9}  \, , \quad 
	b = \frac{5 N_3}{7} \, .
\end{align}
We may rewrite
\begin{align}
	J^{(3,3;6)}(z) &= \sum_m \sum_n \frac{:W_{m} W_{n}:}{z^{m+ n +6}}
	+ \frac{a}{2} \sum_{n} \frac{(n+4)(n+5) U_n}{z^{n+6}} \nonumber \\
	&+ \frac{b}{24 } \sum_{n} \frac{(n+2)(n+3)(n+4)(n+5) L_n}{z^{n+6}} \, ,
	\label{Lambda332}
\end{align}
where the prescription of normal ordering is (see, e.g., (6.144) of \cite{cft}) 
\begin{align}
	{: A B:}_m = \sum_{n \leq - h_A} A_n B_{m -n} + \sum_{n > - h_A} B_{m-n} A_n
	\label{normal}
\end{align}
with $h_A$ as the conformal weight of $A$. 
We then obtain $(C^{(3,3;6)}_{\pm ,0})^2$ with the three point function
\begin{align}
	\langle \mathcal{O}_\pm | J^{(3,3;6)}_0 | \mathcal{O}_\pm \rangle 
	&= \langle \mathcal{O}_\pm | (W_{+2} W_{-2} + W_{+1} W_{-1} + W_0 W_0 + 10 a U_0 + 5 b  L_0 )| \mathcal{O}_\pm \rangle \nonumber \\
	&= \left( \frac{8}{9} u + \frac{1}{14} N_3 h + w^2  \right) \langle \mathcal{O}_\pm | \mathcal{O}_\pm \rangle
\end{align}
and the normalization of higher spin current
\begin{align}
	\langle  J^{(3,3;6)} J^{(3,3;6)} \rangle = 2 \left( - \frac{5 c N_3}{6}\right)^2  \, .
\end{align}

For spin $7$ there is a double trace operator $J^{(3,4;7)} \sim :J^{(3)} J^{(4)}:$  as in \eqref{dtt67}.
As above, we can show that 
\begin{align}
	J^{(3,4;7)} (0) | 0 \rangle  = J^{(3,4;7)}_{-7} | 0 \rangle 
	= \left(W_{-3} U_{-4} + a X_{-7} + b W_{-7} \right) | 0 \rangle
\end{align}
with
\begin{align}
	a =  - \frac{10}{11} \, , \quad 
	b =  - \frac{2}{9} \frac{N_4}{N_3}
\end{align}
is primary. Rewriting
\begin{align}
	J^{(3,4;7)} (z) &= \sum_{m \geq 1 ,n} \frac{:W_{m} U_{n}:}{z^{m+n+7}}
	+ \frac{a}{2} \sum_{n} \frac{(n+5)(n+6) X_n}{z^{n+7}}\nonumber \\
	&  + \frac{b}{ 24} 
	\sum_{n} \frac{(n+3)(n+4)(n+5)(n+6) W_n}{z^{n+7}} \, ,
\end{align}
we find that
\begin{align}
	\langle \mathcal{O}_\pm | J_0^{(3,4;7)} | \mathcal{O}_\pm \rangle
	&= \langle \mathcal{O}_\pm | \left(U_2 W_{-2} + U_1 W_{-1} + U_0 W_0 + 15 a X_0 + 15 b W_0 \right) | \mathcal{O}_\pm \rangle  \nonumber \\ &
	= \left( \frac{15}{11}x - \frac{2}{15}\frac{N_4}{N_3}w + u w \right) \langle \mathcal{O}_\pm | \mathcal{O}_\pm  \rangle \, .
\end{align}
Normalization is given by
\begin{align}
	\langle J^{(3,4;7)} J^{(3,4;7)}  \rangle 
	= \left( - \frac{5 c N_3}{6} \right)\left( - \frac{7 c N_4}{6} \right)  \, .
\end{align}

There are three types as in \eqref{dtt8} for $s=8$, and 
we start from $J^{(4,4;8)} \sim :J^{(4)} J^{(4)}:$.
We assume its form as
\begin{align}
	J^{(4,4;8)} (0)| 0 \rangle  = 
	J^{(4,4;8)}_{- 8} | 0 \rangle 
	= (U_{-4} U_{-4} + a Y_{-8} + b U_{- 8} + d L_{-8} ) | 0 \rangle \, .
\end{align}
The condition $L_{1} J^{(4,4;8)}_{-8} | 0 \rangle = 0$ fixes the constants as
\begin{align}
	a = - \frac{21}{13} \, , \quad b = \frac{42}{11} n_{44} \, , \quad d = \frac{7 N_4}{9} \, .
\end{align}
The operator $J^{(4,4;8)}$ is then obtained as 
\begin{align}
	J^{(4,4;8)} (z) &= \sum_{m,n} \frac{: U_m U_n:}{z^{m+n+8}} + \frac{a}{2} \sum_n
	\frac{(n+6)(n+7) Y_n}{z^{n+8}} + \frac{b}{24} \sum_{n} \frac{(n+4) \cdots (n+7 ) U_n }{z^{n+8}}  \nonumber \\ &+ \frac{d}{6!}\sum_{n} \frac{(n+2) \cdots (n+7) L_n}{z^{n+8}} \, .
\end{align}
Thus we find
\begin{align}
	&\langle \mathcal{O}_\pm | J^{(4,4;8)}_0 | \mathcal{O}_\pm \rangle \nonumber \\
	&= 
	\langle \mathcal{O}_\pm | (U_{3} U_{-3} + U_2 U_{-2 } + U_1 U_{-1} + U_0 U_0 + 21 a Y_0 + 35 b U_0 + 7 d L_0 | \mathcal{O}_\pm \rangle  \\
	& = \left( -\frac{h N_4}{45}+\frac{18 n_{44} u}{11}+u^2+\frac{27 y}{13} \right) \langle \mathcal{O}_\pm | \mathcal{O}_\pm \rangle   \, . \nonumber 
\end{align}
The normalization is
\begin{align}
	\langle J^{(4,4;8)} J^{(4,4;8)} \rangle = 2 \left(  - \frac{7 c N_4}{6} \right)^2  \, .
\end{align}

We then move to $J^{(3,5;8)} \sim :J^{(3)} J^{(5)}:$ in \eqref{dtt8}.
We find
\begin{align}
	J^{(3,5;8)} (0) | 0 \rangle  = 
	J^{(3,5;8)}_{- 8} | 0 \rangle 
	= (W_{-3} X_{-5} + a Y_{-8} + b U_{- 8} ) | 0 \rangle 
\end{align}
with 
\begin{align}
	a = - \frac{10}{13} \, , \quad b = - \frac{15}{154} \frac{N_5}{N_4} 
\end{align}
is primary.
With this expression, we compute
\begin{align}
	\langle \mathcal{O}_\pm | J^{(3,5;8)}_0 | \mathcal{O}_\pm \rangle& = 
	\langle \mathcal{O}_\pm | (X_{2} W_{-2} + X_1 W_{-1} + X_0 W_{0} + 21 a Y_0 + 35 b U_0 | \mathcal{O}_\pm \rangle  \\
	& =\left(  -\frac{15 N_5 u}{77 N_4}+w x+\frac{24 y}{13} \right) \langle \mathcal{O}_\pm | \mathcal{O}_\pm \rangle   
\end{align}
and
\begin{align}
	\langle \Lambda^{(3,5)} \Lambda^{(3,5)} \rangle =  \left(  - \frac{5 c N_3}{6} \right) \left(  - \frac{9 c N_5}{6} \right)  \, .
\end{align}

For $J^{(3,3;8)} \sim : J^{(3)} \partial^2 J^{(3)}:$ in \eqref{dtt8}, we define
\begin{align}
	J^{(3,3;8)}_{- 8} | 0 \rangle 
	= (W_{-5} W_{-3} + a W_{-4} W_{-4} + b U_{- 8} + d L_{-8} ) | 0 \rangle \, ,
\end{align}
where the condition $L_{1} J^{(3,3;8)}_{-8} | 0 \rangle = 0$ leads to
\begin{align}
	a = - \frac{7}{12} \, , \quad b = \frac{7}{11}  \, , \quad d = - \frac{35 N_3}{36} \, .
\end{align}
Using 
\begin{align}
	J^{(3,3;8)}(z) &= \frac12 \sum_{m,n} \frac{:(m+3)(m+4) W_m W_n:}{z^{m+n+8}} + a \sum_{m,n}
	\frac{: (m+3) W_m (n+3) W_n :}{z^{m + n+8}} \nonumber \\ & + \frac{b}{24} \sum_{n} \frac{(n+4) \cdots (n+7 ) U_n }{z^{n+8}} + \frac{d}{6!} \sum_{n} \frac{(n+2) \cdots (n+7) L_n}{z^{n+8}} \, ,
\end{align}
 we find
\begin{align}
	\langle \mathcal{O}_\pm | J^{(3,3;8)}_0 | \mathcal{O}_\pm \rangle &= 
	\langle \mathcal{O}_\pm | \tfrac12 ( 2 W_2 W_{-2 } + 6 W_1 W_{-1} + 12 W_0 W_0 )  \\
	&+ a (5 W_2 W_{-2 } + 8 W_1 W_{-1} + 9 W_0 W_0 )  + 21 b Y_0 + 35 d U_0  | \mathcal{O}_\pm \rangle \nonumber \\
	& = \left( \frac{h N_3}{36}+\frac{3 u}{11}+\frac{3 w^2}{4} \right) \langle \mathcal{O}_\pm |  \mathcal{O}_\pm \rangle \, . \nonumber 
\end{align}
Here we have applied the normal ordering prescription as in \eqref{normal}.
For instance, we may set 
\begin{align}
	A_m = (m+3) (m+4) W_m \, , \quad B_n = W_n , \quad h_A = 5 \, .
\end{align}
The normalization is
\begin{align}
	\langle J^{(3,3;8)} J^{(3,3;8)}\rangle = 
	\left(  - \frac{5 c N_3}{6} \right)  \left(  - \frac{35 c N_3}{2} \right) + 
	2  a^2 \left(  - 5 c N_3 \right)^2  \, .
\end{align}

There could be another spin 8 current of double trace type as
$J^{(3,4;8)} \sim :J^{(3)} \partial J^{(4)}:$. We can see that 
\begin{align}
J^{(3,4;8)} (0) | 0 \rangle = 
J^{(3,4;8)}_{- 8} | 0 \rangle 
= (W_{-3} U_{-5} + a W_{-4} U_{-4} + b X_{- 8} + d W_{-8} ) | 0 \rangle 
\end{align}
is primary for 
\begin{align}
a = - \frac{4}{3} \, , \quad b = - \frac{5}{3}  \, , \quad d = - \frac{4 N_4}{5 N_3} \, .
\end{align}
Since $J^{(3,4;8)} (z)$ is given by
\begin{align}
J^{(3,4;8)} (z) &= -  \sum_{m,n} \frac{: W_m (n+4) U_n:}{z^{m+n+8}} - a \sum_{m,n}
\frac{: (m+3) W_m U_n :}{z^{m + n+8}} \nonumber \\ & - \frac{b}{6} \sum_{n} \frac{(n+5) (n+6) (n+7 ) X_n }{z^{n+8}} - \frac{d}{5!} \sum_{n} \frac{(n+3) \cdots (n+7) W_n}{z^{n+8}} \, ,
\end{align}
we find
\begin{align}
\langle \mathcal{O}_\pm | J^{(3,4;8)}_0 | \mathcal{O}_\pm \rangle &= 
\langle \mathcal{O}_\pm | -  ( 6 U_2 W_{-2 } + 5 U_1 W_{-1} + 4 U_0 W_0 )  \\
& - a ( U_2 W_{-2 } + 2 U_1 W_{-1} + 3 U_0 W_0 )  - 35 b X_0 - 21 d  W_0  | \mathcal{O}_\pm \rangle = 0 \, , \nonumber 
\end{align}
which means that there is no contribution from $J^{(3,4;8)} $.


\providecommand{\href}[2]{#2}\begingroup\raggedright\endgroup

\end{document}